\documentclass[12pt]{article}
\usepackage{latexsym,graphicx,amssymb,color,amsmath}
\usepackage{lscape}
\usepackage{epsfig}
\date{}
\newcommand{\be}{\begin{equation}}
\newcommand{\ee}{\end{equation}}
\newcommand{\ba}{\begin{eqnarray}}
\newcommand{\ea}{\end{eqnarray}}
\newcommand{\hf}{\frac{1}{2}}
\renewcommand{\theequation}{\thesection.\arabic{equation}} 
\begin{document}
\vspace*{-2cm}
\begin{flushright}
DCPT/07/57
\end{flushright}

\vspace{2cm}

\begin{center}
{\Large {\bf Dynamical Implications of Viral Tiling Theory}}\\ 
\vspace{1cm} {\large  \bf K. M. 
ElSawy\,\footnote{\noindent E-mail: {\tt ke112@york.ac.uk}}, A.\
Taormina\,\footnote{\noindent E-mail: {\tt anne.taormina@durham.ac.uk}},
R.\ Twarock\,\footnote{\noindent E-mail: 
{\tt rt507@york.ac.uk}} and L. Vaughan\,${}^{4}$}\\
\vspace{0.3cm} {${}^{1,3}$}\em Department of Mathematics and Department of Biology\\ University of York\\
York YO10 5DD, U.K.\\
 \vspace{0.3cm} {${}^{2,4}$\em \it Department of Mathematical
Sciences\\ University of Durham\\ Durham DH1 3LE, U.K.}\\ 
\end{center}

\begin{abstract}
\noindent The Caspar-Klug classification of viruses whose protein shell, called viral capsid, exhibits icosahedral symmetry, has recently been extended to incorporate viruses whose capsid proteins are exclusively organised in pentamers. The approach, named `Viral Tiling Theory', is inspired by the theory of quasicrystals, where aperiodic Penrose tilings enjoy 5-fold and 10-fold local symmetries. This paper analyzes the extent to which this classification approach informs dynamical properties of the viral capsids, in particular the pattern of Raman active modes of vibrations, which can be observed experimentally.
\end{abstract}
\vskip 1cm
\section{Introduction}

Viruses are dynamical particles which can undergo vibrational movements. The study of normal modes of vibration of viral capsids may shed light on conformational changes of viruses, help understand the release of genetic material \cite{Sherman} in the viral replication process\footnote{Observations indicate that the RNA release in the Tomato Bushy Stunt Virus happens through tiny apertures located at local symmetry axes of the icosahedral capsid,  whose formation is consistent with the effect of certain normal modes of vibration  \cite{Orlova}.} and reveal how chemical bound compounds affect the flexibility of the capsid
and hence inhibit the infectiousness of the virus.  To analyze vibrational patterns in viral capsids, one requires knowledge of the capsid's structure and its energy function. In most cases, a classical mechanics treatment for viruses kept at a temperature close to $0^o$ K, governed by a harmonic potential, yields a reasonable picture of any viral motion, which can be exactly expressed as a superposition of normal modes. Among those, the most relevant in applications are the lowest frequency modes, which correspond to delocalized motions of a large number of atoms.

Several detailed analyses of normal modes of vibrations have appeared recently in the literature \cite{TamaBrooks, VanVlijmen},  where the capsid's structure is considered at various stages of coarse-graining, and where the potential mimics the effect of springs in-between the capsid's constituents in a given coarse-grained approximation \cite{Tirion}.
These calculations require several computational prowesses, as more accurate descriptions of the capsid's structure involve a rapidly increasing number of degrees of freedom. Group theoretical methods have long been an important ally in vibrational analyses \cite{Simonson}, particularly in the calculation of the normal modes of vibration of fullerenes \cite{Weeks, James, Nakamoto}. They enhance the performance and have recently been used in studies of vibration patterns for small viruses at the atomic level, with a basis set including all the internal dihedral angles of the system considered, except
for peptide bonds which are assumed to be rigid \cite{VanVlijmen}. These studies however are very much done case by case, and do not capture potential patterns of vibrations within certain classes of viruses. 

The present note investigates whether Viral Tiling Theory, a recently proposed model
for icosahedral viral capsids which solves a classification puzzle in the Caspar-Klug nomenclature \cite{Twarock}, provides a new qualitative insight in the dynamics of viruses. In particular, we ask whether there is a correlation between the vibrational patterns of viruses with a given number  of coat proteins and their viral tiling.
The following analysis is mainly qualitative and informs on the group theoretical properties of the normal modes of vibration. These properties  enable to identify  which normal modes can be detected by Raman and infrared spectroscopy. The calculation of the relevant frequencies of vibration requires further techniques which we postpone to a future publication \cite{Peeters Taormina}.

The paper is organised as follows. In Section 2, we present the tilings relevant to the description of a variety of viruses with triangulation numbers $T=3$ and $T=7$, together with their maximally symmetric decorations. We then confront these ideally decorated tiles with experimental data for Bacteriophages MS2 and HK97, as well as for Tomato Bushy Stunt Virus and Simian Virus 40 in Section 3. We argue that MS2 and SV40 capsids exhibit a centre of inversion in good approximation. In Section 4, we use the group theory   
underlying the icosahedral symmetry of the viral capsids considered to obtain qualitative information on their normal modes of vibration, and we offer some conclusions in the last section.
\section{Ideal tilings of icosahedral viral capsids}

The mathematical structure underpinning the symmetry of icosahedral capsids  is the non-crystallographic Coxeter group $H_3$ (also labelled ${\cal I}_h$ in the science literature)\cite{Humphreys}.  It contains 120 elements and is generated by a 2-fold rotation $g_2$ and a 3-fold rotation $g_3$, chosen according to the rules in Appendix A, as well as by the inversion transformation $g_0$ which maps each 3-dimensional point of coordinates $(x,y,z)$ to $-(x,y,z)$. The sixty proper rotations generated by $g_2$ and $g_3$ form the subgroup ${\cal I}$, which is the relevant group for the study of vibrational patterns whenever the viral capsid does not possess a centre of inversion. By this we mean that the distribution of  the $N$ capsid constituents $C_i, i=1,..,N$, each  being represented by a vector whose origin coincides with the centre of the capsid and whose components are $(x_i,y_i,z_i)$,  is not invariant under the inversion operation which maps $(x_i,y_i,z_i)$ to $-(x_i,y_i,z_i)$. Although experimental data do not support the existence of capsids with strict centre of inversion, we argue in Section 3 that some have it in very good approximation, and therefore their normal modes of vibration can be analysed  with the help of the full icosahedral group rather than its subgroup ${\cal I}$.

The affinization of $H_3$ has been recently constructed in \cite{Patera} and can be used to determine the locations of all global and local symmetry axes of viral capsids compatible with icosahedral symmetry
\cite{TwarockKeef}. Once these axes have been identified from group theoretical considerations for a viral capsid of given triangulation number $T$ , it is easy to design spherical tiles with appropriate decorations \footnote{Decorations are dots on the tiles indicating schematically the locations of the capsid proteins.}  which pave the capsid and keep track of the cluster distribution of proteins around the symmetry axes of the capsid, while also encoding the bond structure between those proteins.

There exist several ways to decorate the prototiles of a given tiling with dots representing individual capsid proteins, and still capture their symmetry properties under the global 5-,\,3- and 2-fold rotations of the icosahedron. However, some decorating patterns are distinguished inasmuch as they correspond to capsid protein distributions with maximal symmetry. By this we mean that the distribution of dots on each prototile of such ideal tilings exhibits the highest symmetry compatible with the tile shape.  In what follows, we describe ideal tilings for two types of $T=3$ and $T=7$ viral capsids, as they will be used as references when considering the experimental data for viruses representatives of these triangulation numbers. 

Our choice of $T=3$ and $T=7$ capsids is motivated by our wish  to gain information on whether or not qualitative differences in vibrational patterns are rooted in the nature of the tiling considered (at fixed capsid size) and are independent of the capsid's size.

\subsection{$T=3$ tilings}

The $T=3$  icosahedral capsids  come in three mathematical species, of which two can be used to model viruses observed in vivo. The corresponding tessellations are the Caspar-Klug (CK) tiling, with triangular prototiles encoding trimer interactions, the rhomb tiling with prototiles in the shape of rhombs representing dimer interactions, and a kite tiling whose building blocks encode trimer interactions as in the CK case. We shall not discuss this latter case any further as we are not aware of any $T=3$ virus which could be modelled in this way.

\begin{figure}[ht]
\begin{center}
\raisebox{-1.3cm}{\includegraphics[width=10.1cm,keepaspectratio]{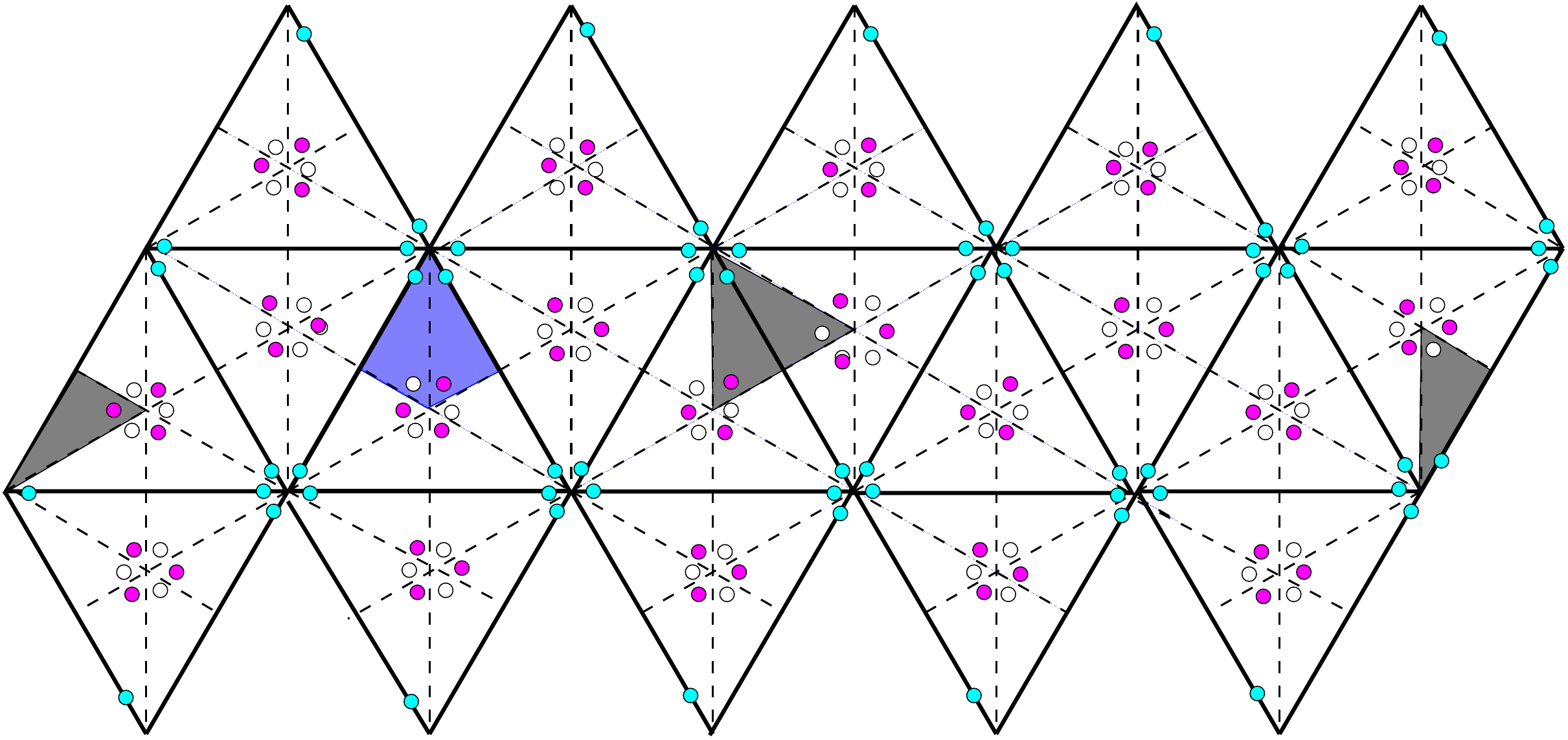}}
\end{center}
\caption{{\em The ideal Caspar-Klug tiling for the $T=3$ viral capsid with the location of proteins in chains A, B and C marked with purple, cyan and white  dots respectively . The grey shaded triangular prototiles highlight trimer interactions between capsid proteins, while the blue shaded region corresponds to the fundamental domain of the  proper rotation subgroup of the full icosahedral group $H_3$ . }}
\label{fig:icosaidealcktiling}
\end{figure}

\begin{figure}[ht]
\begin{center}
\raisebox{-1.3cm}{\includegraphics[width=10.1cm,keepaspectratio]{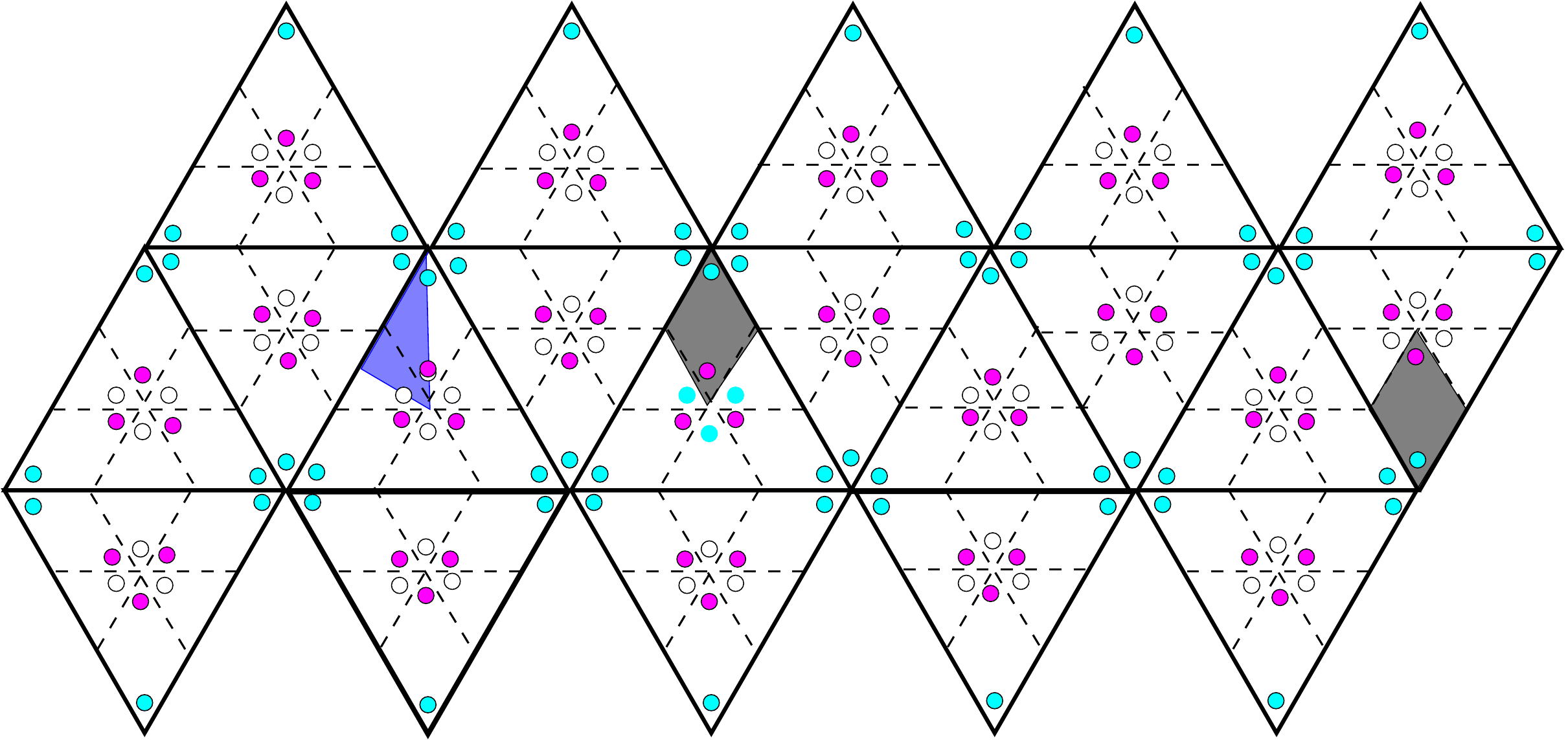}}
\end{center}
\caption{{\em The ideal rhomb tiling for the $T=3$ viral capsid with the location of proteins  in chains A, B and C marked with purple,  cyan and white dots respectively.  The grey shaded rhombic prototiles highlight dimer interactions between capsid proteins, while the blue shaded region corresponds to the fundamental domain of the full icosahedral group $H_3$ . }}
\label{fig:icosaidealrhombictiling}
\end{figure}

The capsid proteins are organised in pentamers (clusters of five)  about the 12 global 5-fold symmetry axes, and hexamers (clusters of six) about 20 local 6-fold symmetry axes, which coincide with the 20 global 3-fold symmetry axes of the icosahedron. The total number of proteins is therefore $180=12 \times 5 + 20 \times 6$. The ideal tilings shown on a planar representation of the icosahedron in Fig.~\ref{fig:icosaidealcktiling} and Fig.~\ref{fig:icosaidealrhombictiling} take into account the fact that $T=3$ capsid proteins belong to three different chains \footnote{Polypeptide chains representing capsid proteins.} - symbolised by three different colours - but their location on the prototiles does not represent certain aspects of the experimental data, such as the exact locations of the centres of mass of the capsid proteins, as will be discussed in the following section.

It is interesting to note that the CK $T=3$ ideal tiling does not exhibit a centre of inversion when considering the action of the icosahedral group on individual proteins, as is most obvious in the structure of the fundamental domain of the subgroup ${\cal I}$ shaded in blue on the figures: a centre of inversion does exist if all dots in the kite-shape fundamental domain are invariant under a reflection through  the unique axis of symmetry of the kite. 
This fails to happen in the CK tiling if the  white and purple dots represent proteins of different types and masses, which is usually the case.  For example, the masses of the three proteins of the Tomato Bushy Stunt Virus are (in atomic mass units) 28149.5, 28067.4 and 31311.5 respectively. There is hence a non-negligible difference of 3244.1 atomic mass units between the two proteins constituting the hexamer. 
The situation is different however for the rhomb tiling  of Fig.~\ref{fig:icosaidealrhombictiling}, where a centre of inversion of the ideal tiling is present, and the fundamental domain in this case is reduced to half the kite of Fig.~\ref{fig:icosaidealcktiling}.
cut through the symmetry axis.

\subsection{$T=7$ tilings}
\begin{figure}[ht]
\begin{center}
\raisebox{-1.3cm}{\includegraphics[width=13.1cm,keepaspectratio]{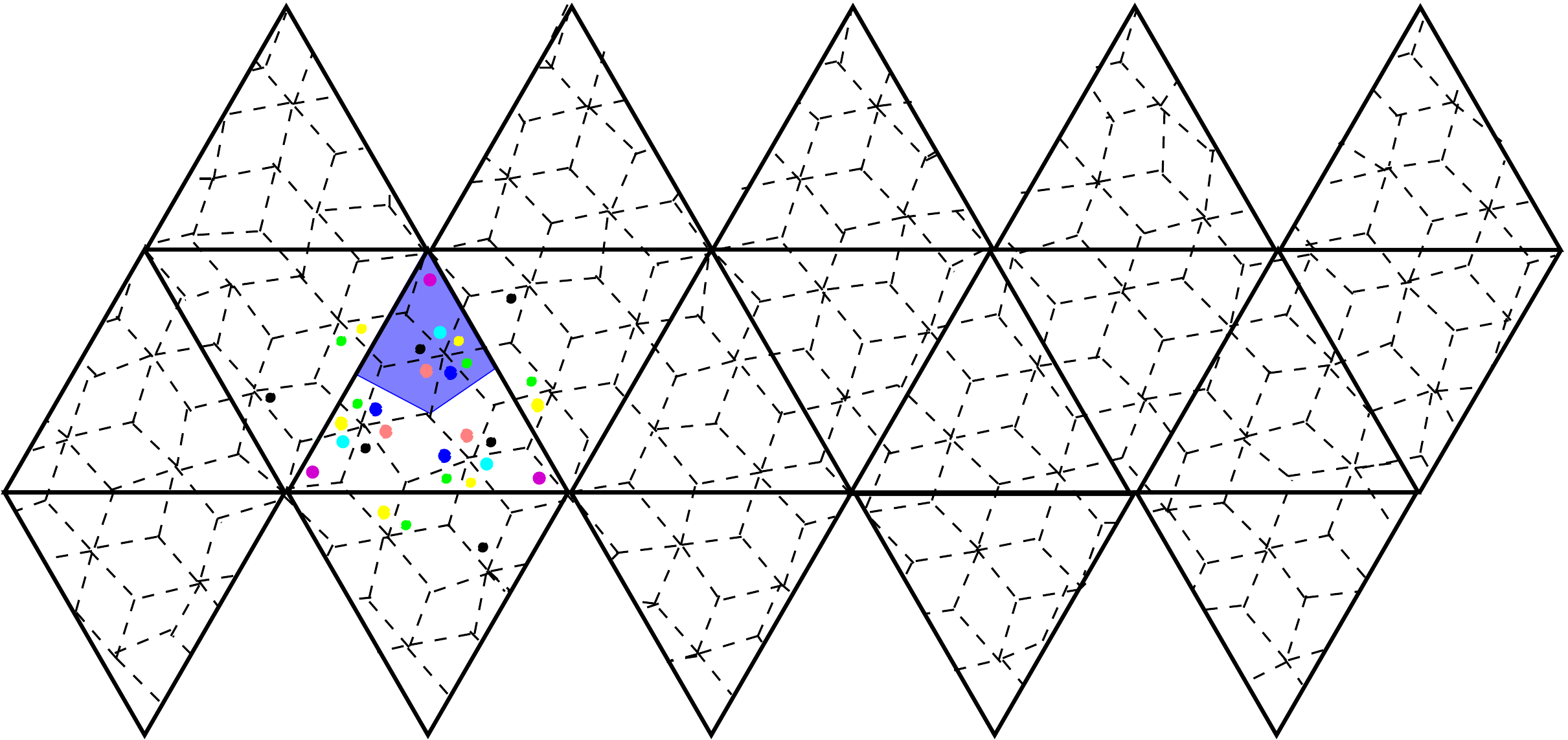}}
\end{center}
\caption{{\em The ideal rhomb tiling for the $T=7\ell$ viral capsid with the location of proteins  in chains A, B, C, D, E, F and G marked with purple, cyan, white, green, pink, magenta  and yellow dots respectively.  The  blue shaded region corresponds to the fundamental domain of the  proper rotation subgroup of the full icosahedral group $H_3$ . }}
\label{fig:icosahk97}
\end{figure}

We now construct two ideal tilings for $T=7$ capsids. The first one has left chirality
($T=7\ell$) and accommodates 420 capsid proteins, with four types of dimer interactions modelled by rhomb prototiles as shown in Fig.~\ref{fig:icosahk97}. Note that the fundamental domain shaded in blue indicates this tiling does not have a centre of inversion, since the distribution of dots within it is not symmetrical under a reflection through the symmetry axis of the kite-shaped domain.
\begin{figure}[ht!]
\begin{center}
\raisebox{-1.3cm}{\includegraphics[width=13.1cm,keepaspectratio]{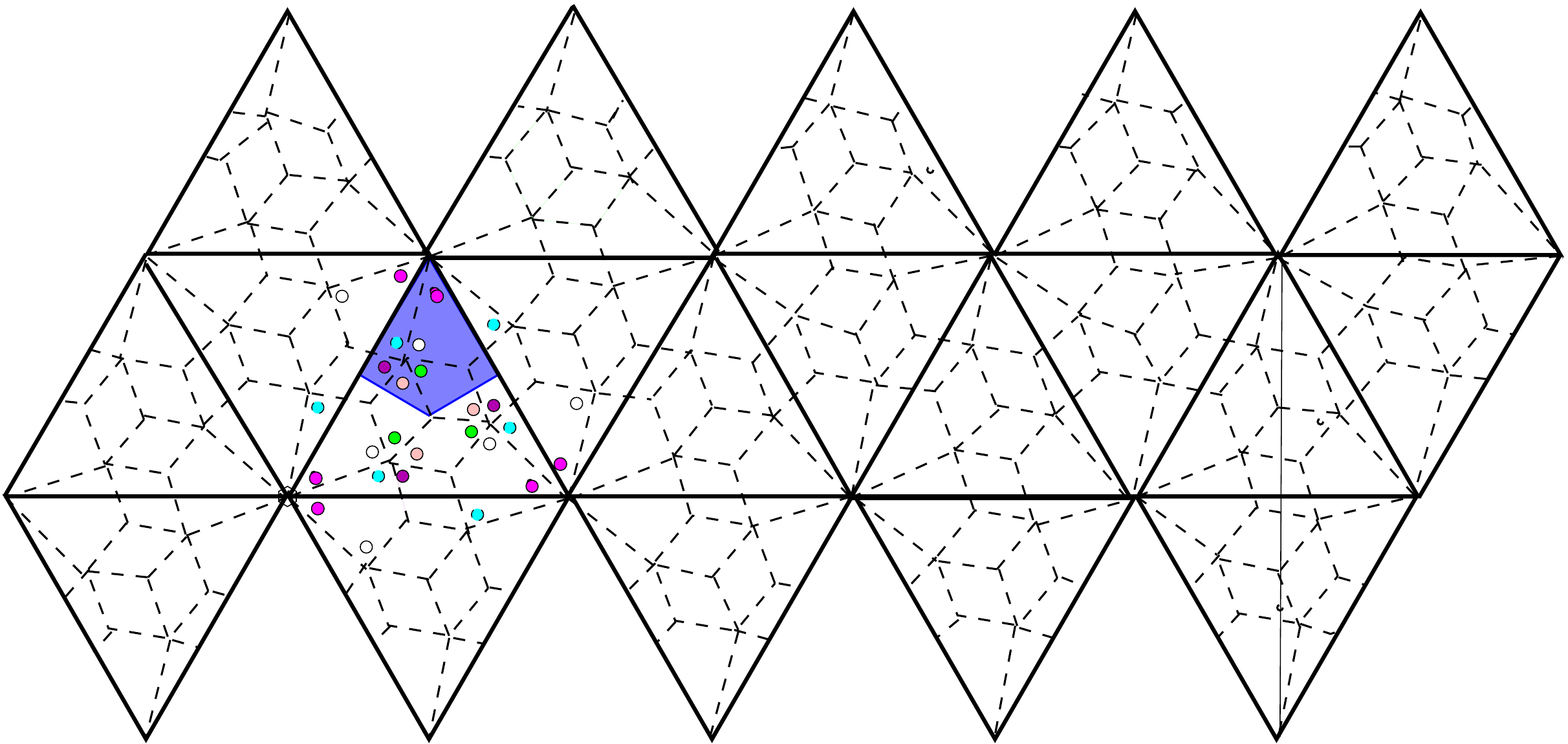}}
\end{center}
\caption{{\em The ideal rhomb and kite tiling for the $T=7d$ all-pentamer capsid with the location of proteins  in chains A, B, C, D, E, and F  marked with purple, cyan, white, green, pink and magenta  dots respectively.  The  blue shaded region corresponds to the fundamental domain of the  proper rotation subgroup of the full icosahedral group $H_3$. }}
\label{fig:sv40bis}
\end{figure}

The second capsid is of right chirality ($T=7d$) but only accommodates 360 capsid proteins organised in pentamers through two types of prototiles, namely rhombs encoding two types of dimer interactions, and kites, encoding trimer interactions. Fig.~\ref{fig:sv40bis} clearly shows that this ideal capsid does not have a centre of inversion.
We now confront our ideal tilings with experimental data for $T=3$ and $T=7$ viruses
and draw conclusions on actual capsid symmetries.

\section{Experimental data and viral capsid symmetries}

In order to extract qualitative features of vibrational patterns from viral capsids, we first restrict ourselves to a coarse-grained approximation where each capsid protein is replaced by a point mass whose location coincides with the centre of mass of the protein considered. This centre of mass is calculated by taking into consideration all 
crystallographically identified 
atoms of the protein, according to data stored in the Protein Data Bank or equivalently the VIPER website. 

We then quantify the deviation between the capsid protein distribution obtained in this approximation and the distribution obtained from it by inversion. At the light of this information, we finally discuss, for each virus considered, whether the capsid exhibits a centre of inversion in good approximation. 

\subsection{Bacteriophage MS2 and Tomato Bushy Stunt Virus}
\begin{figure}[ht]
\begin{center}
(a)\includegraphics[width=4.7cm,keepaspectratio]{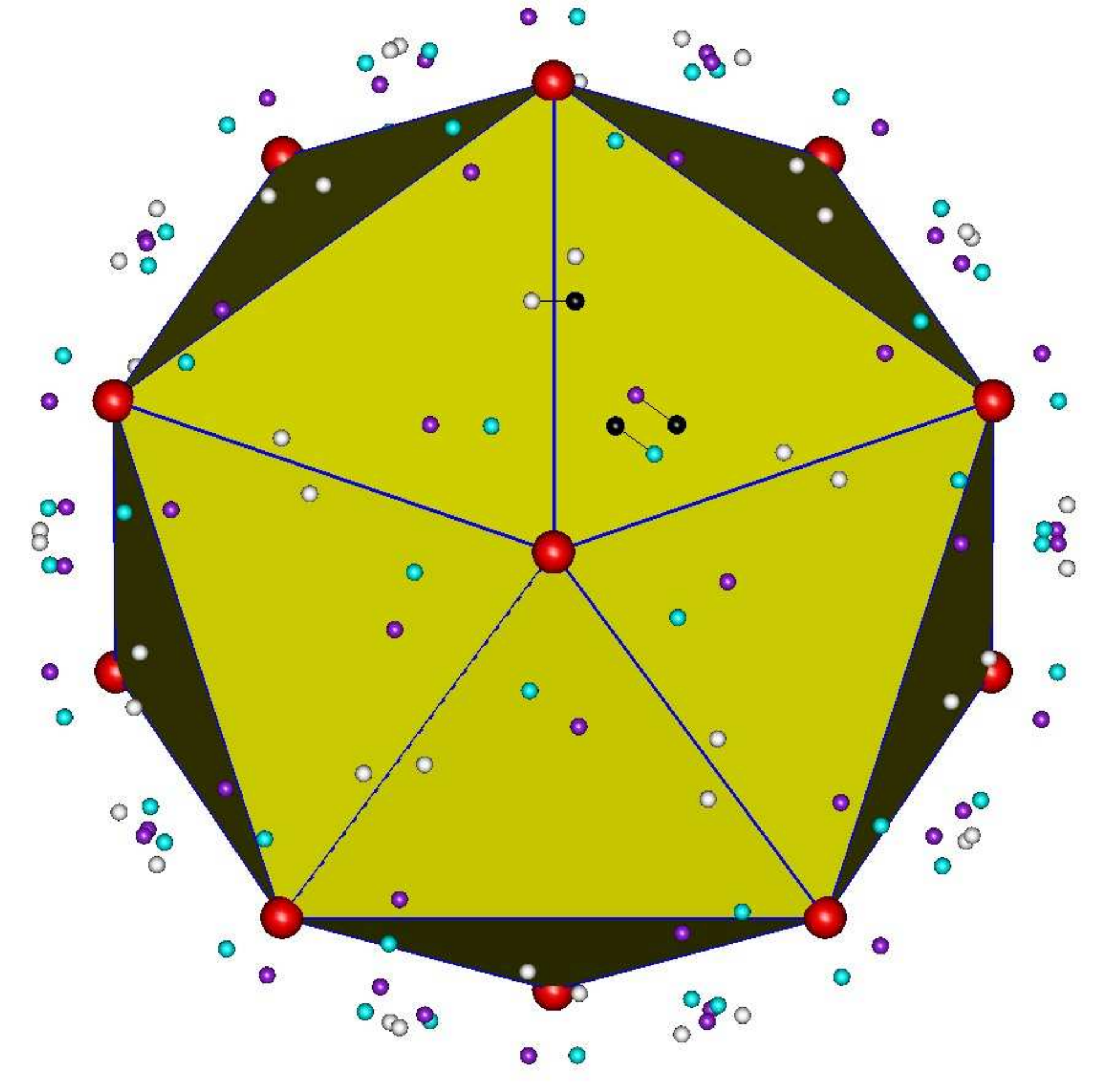}
(b)\includegraphics[width=5.2cm,keepaspectratio]{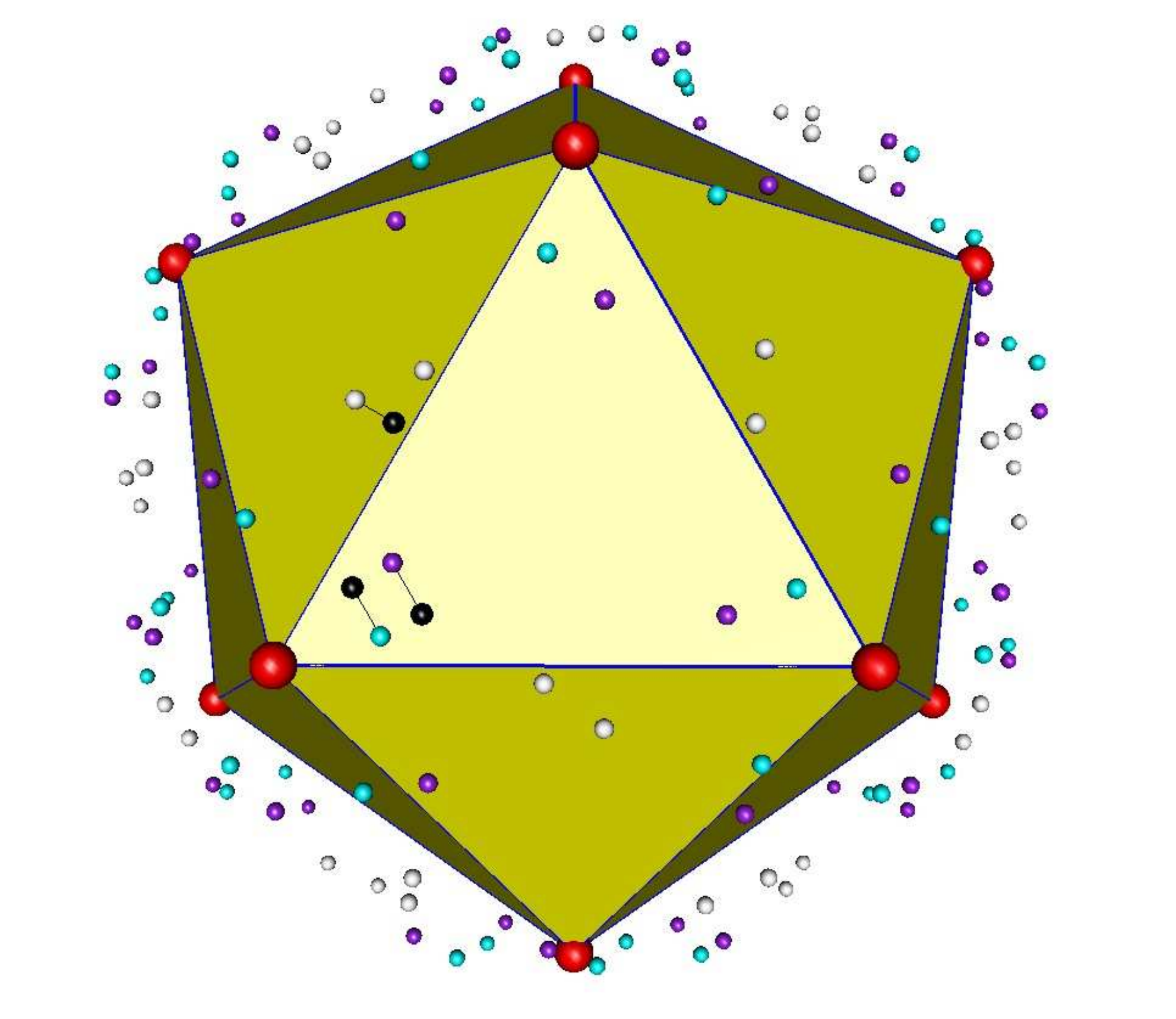}
(c)\includegraphics[width=4cm,keepaspectratio]{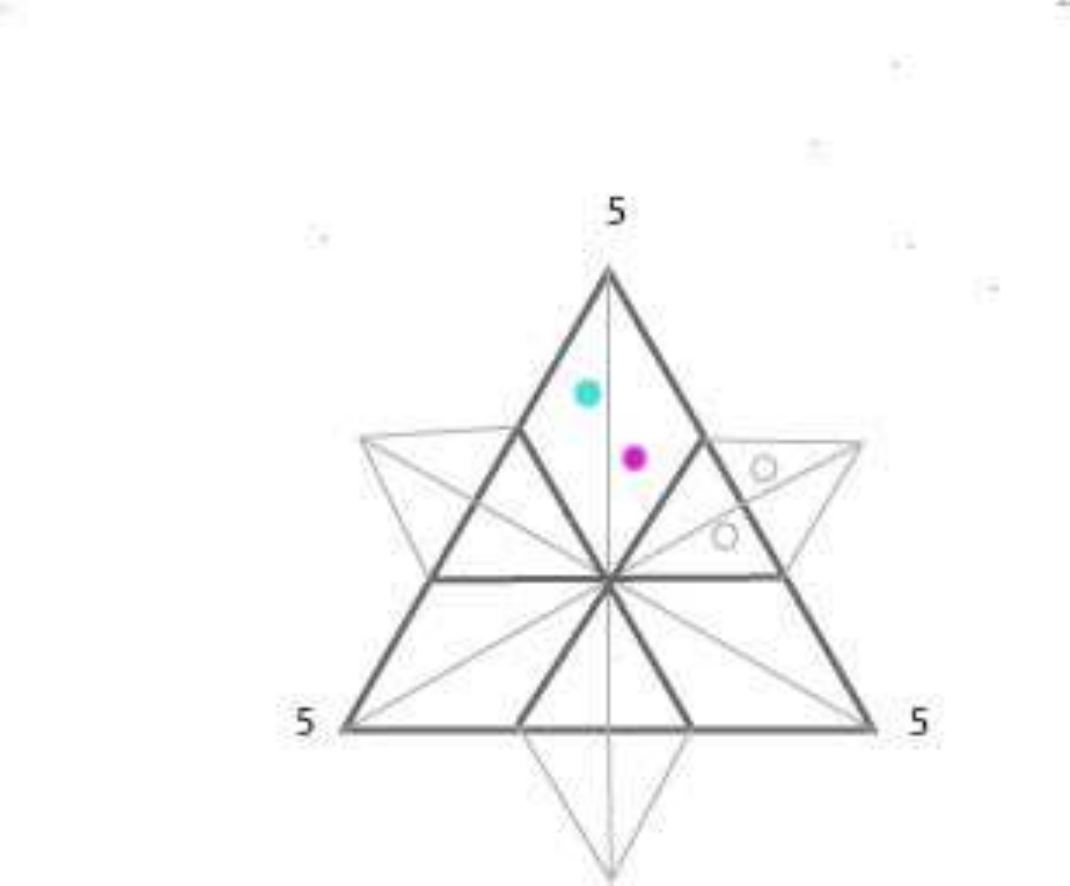}
\end{center}
\caption{\em{The calculated centres of mass for the capsid proteins in chains A, B and C of Bacteriophage MS2 are represented by big dots in purple, cyan and white respectively,  while the small black dots correspond to the distribution of centres of mass after inversion. Views from above a 5-fold (a) and a 3-fold (b) symmetry axes are provided; (c) Rhomb tiling decorations induced by experimental data.}}
\label{MS2}
\end{figure}

We start by considering the case of MS2, a $T=3$ capsid with 180 capsid proteins organised in a first approximation according to a rhomb tiling. Based on the .pdb (or .vdb) file with ID 2ms2\,\footnote{See http://viperdb.scripps.edu/index.php or http://www.rcsb.org/pdb/home/home.do}, we represent in Fig.~\ref{MS2} the centre of mass of the A, B and C chains in purple, cyan and white respectively.  

We use this information to derive the qualitative locations of these vertices with respect to a  $T=3$ rhomb tiling. The result is shown for an icosahedral triangle in Fig.~\ref{MS2}(c). In Fig.~\ref{MS2}(a) and (b), we show in black some of the vertices obtained by inversion from vertices located on the far side of the capsid. The distances between these black vertices and the vertices to which they are connected by a black line correspond to twice the deviation from inversion symmetry for these vertices. 
The deviation from inversion symmetry can be determined for each chain, and we obtain the figures given in Table~\ref{MS2Table}. Although Bacteriophage MS2 is determined at a resolution of 2.8 $\AA$, we argue that the small deviations in Table~\ref{MS2Table} are consistent with the assumption that the viral capsid has an approximate centre of inversion. 

We thus determine symmetry-corrected versions of the tiling by averaging the position of each protein with that of an inverted one (of the same chain) in its immediate neighbourhood. 

\begin{figure}[ht]
\begin{center}
(a)\includegraphics[width=4.9cm,keepaspectratio]{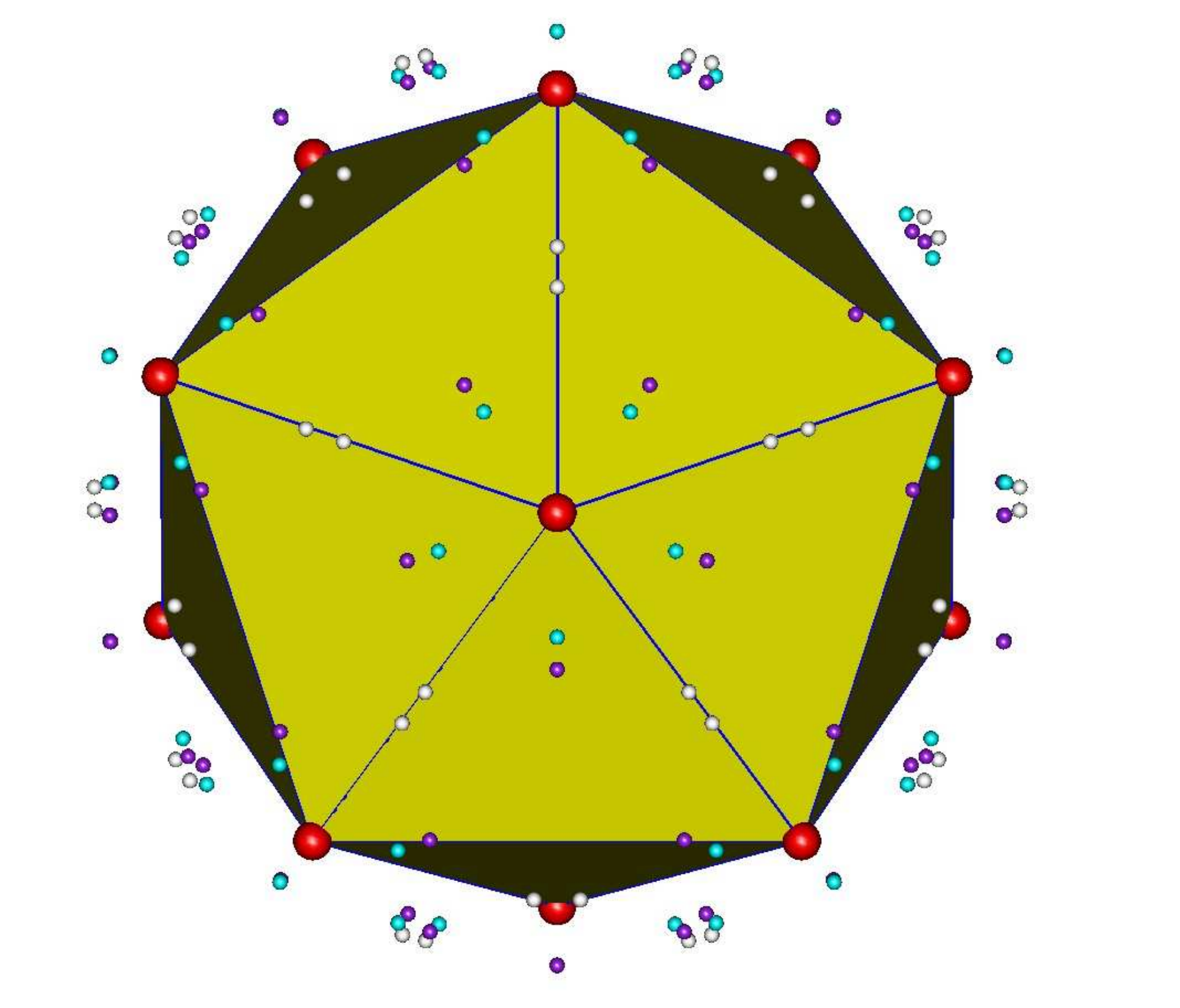}
(b)\includegraphics[width=5cm,keepaspectratio]{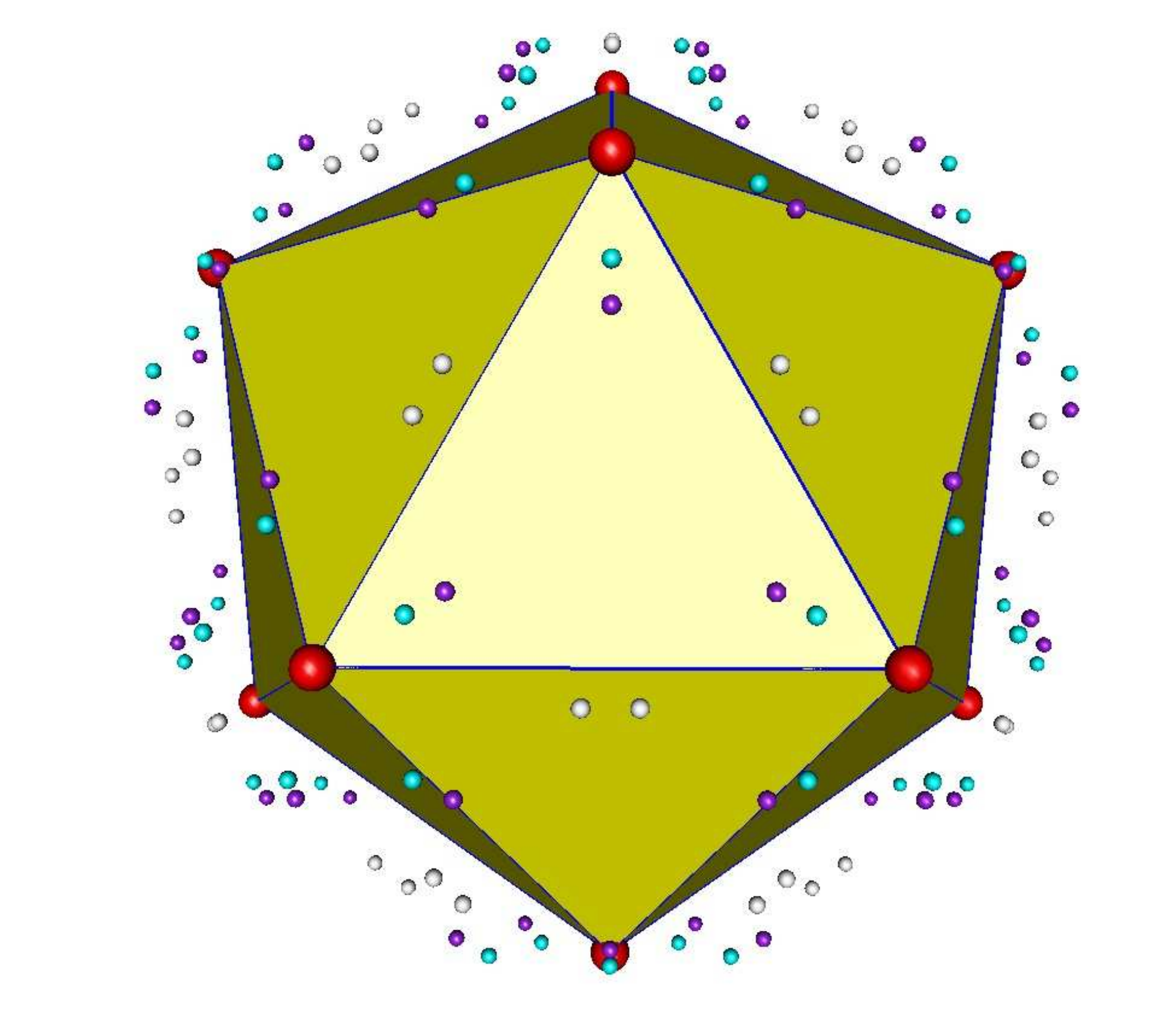}
(c)\includegraphics[width=4.0cm,keepaspectratio]{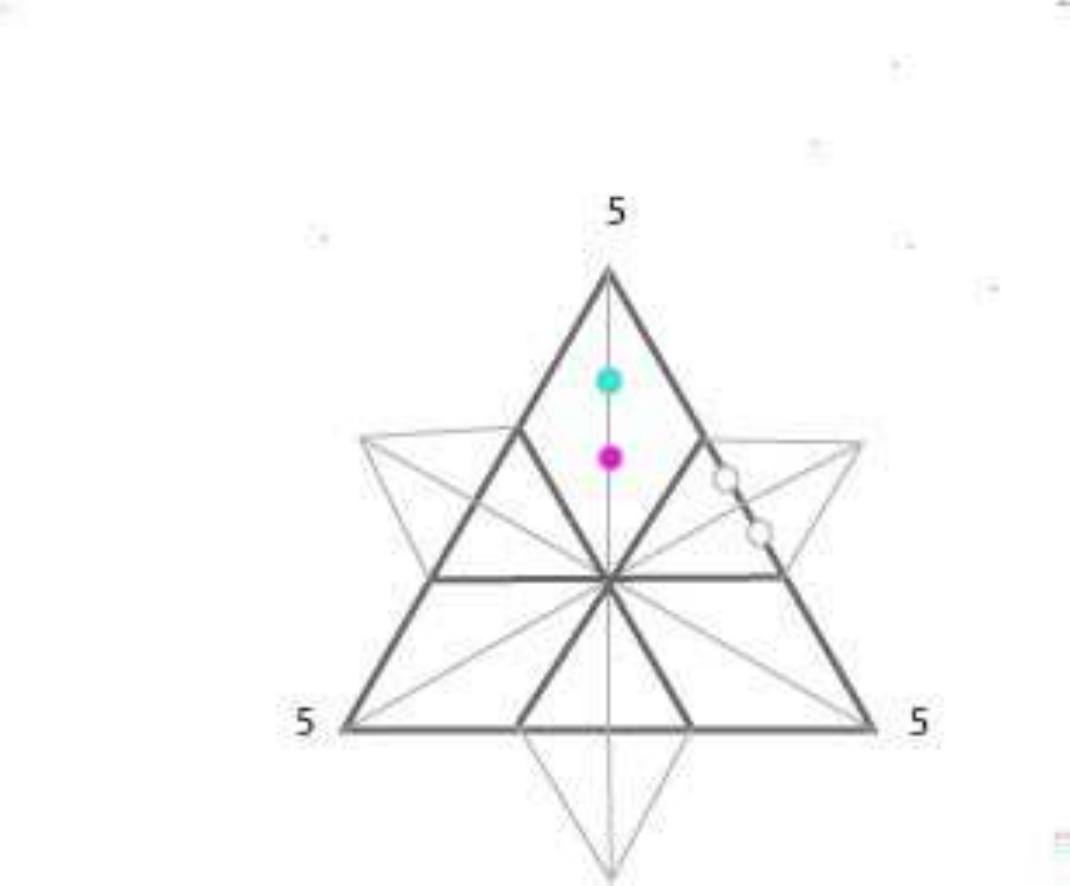}
\end{center}
\caption{\em{Symmetry-corrected MS2 capsids of Type 1 (a)  along a 5-fold axis, (b)  along a 3-fold axis, (c) corresponding tiling decorations.}}
\label{MS21}
\end{figure}

\begin{figure}[ht]
\begin{center}
(a)\includegraphics[width=5.2cm,keepaspectratio]{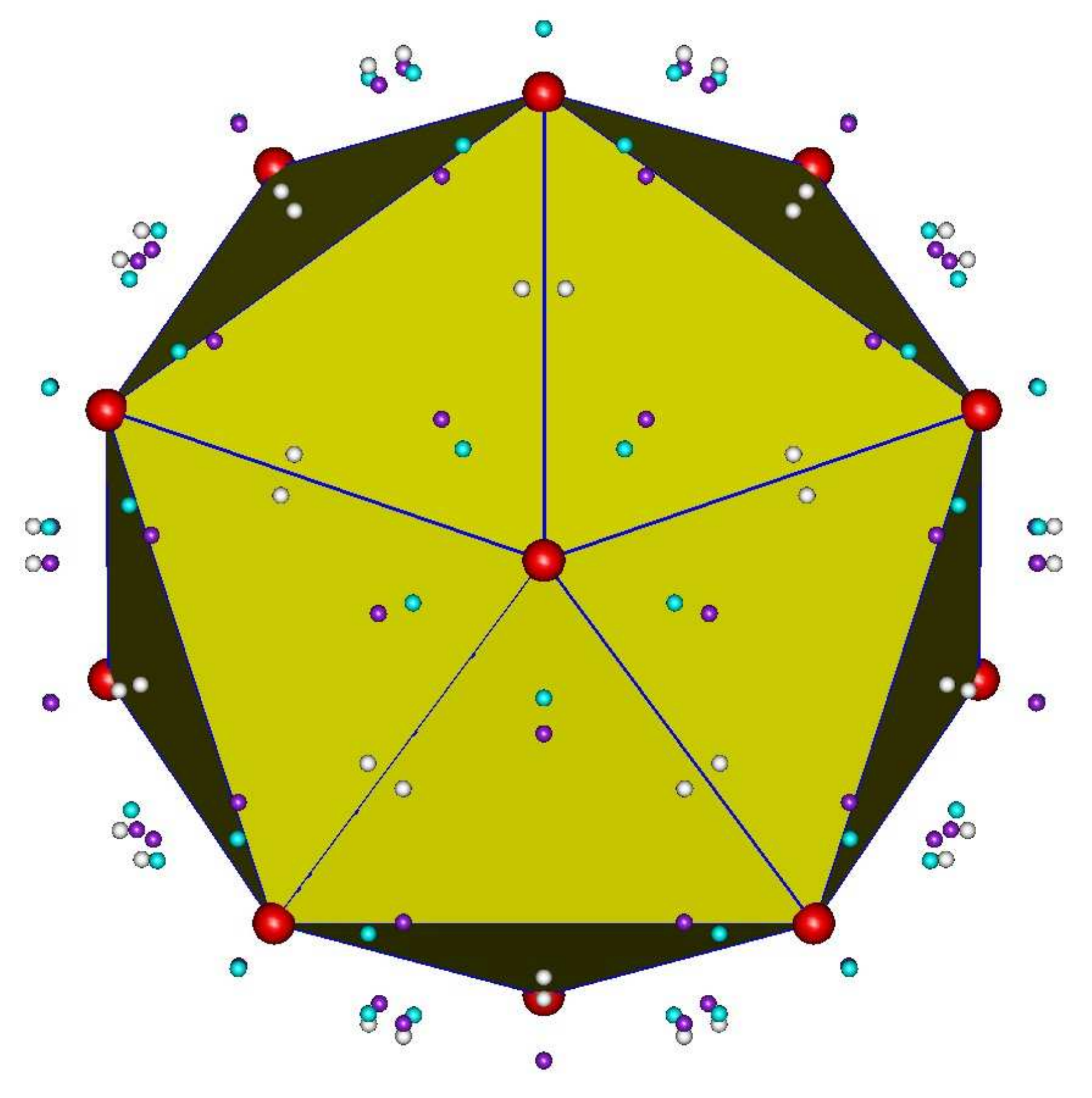}
(b)\includegraphics[width=4.7cm,keepaspectratio]{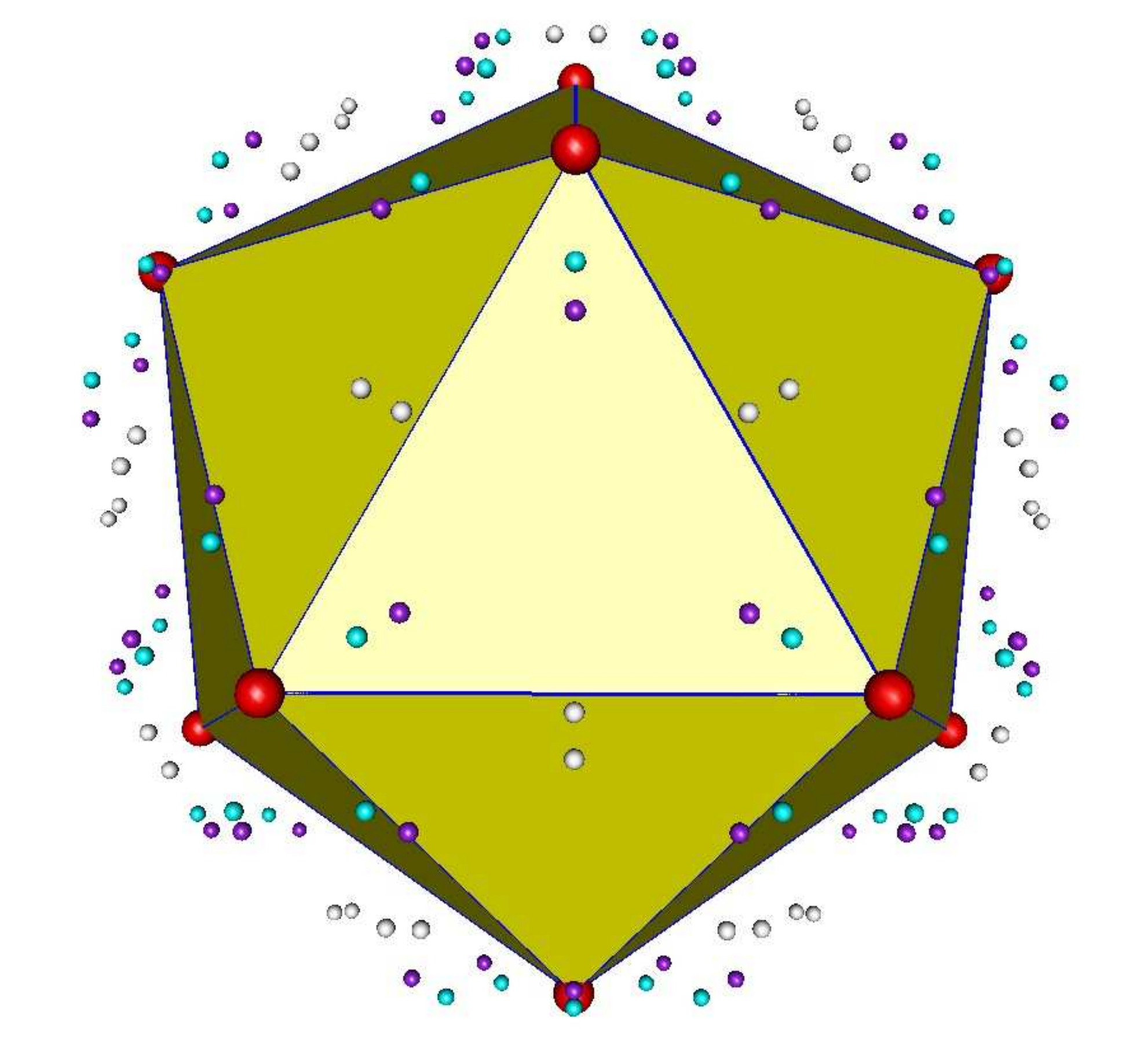}
(c)\includegraphics[width=4.0cm,keepaspectratio]{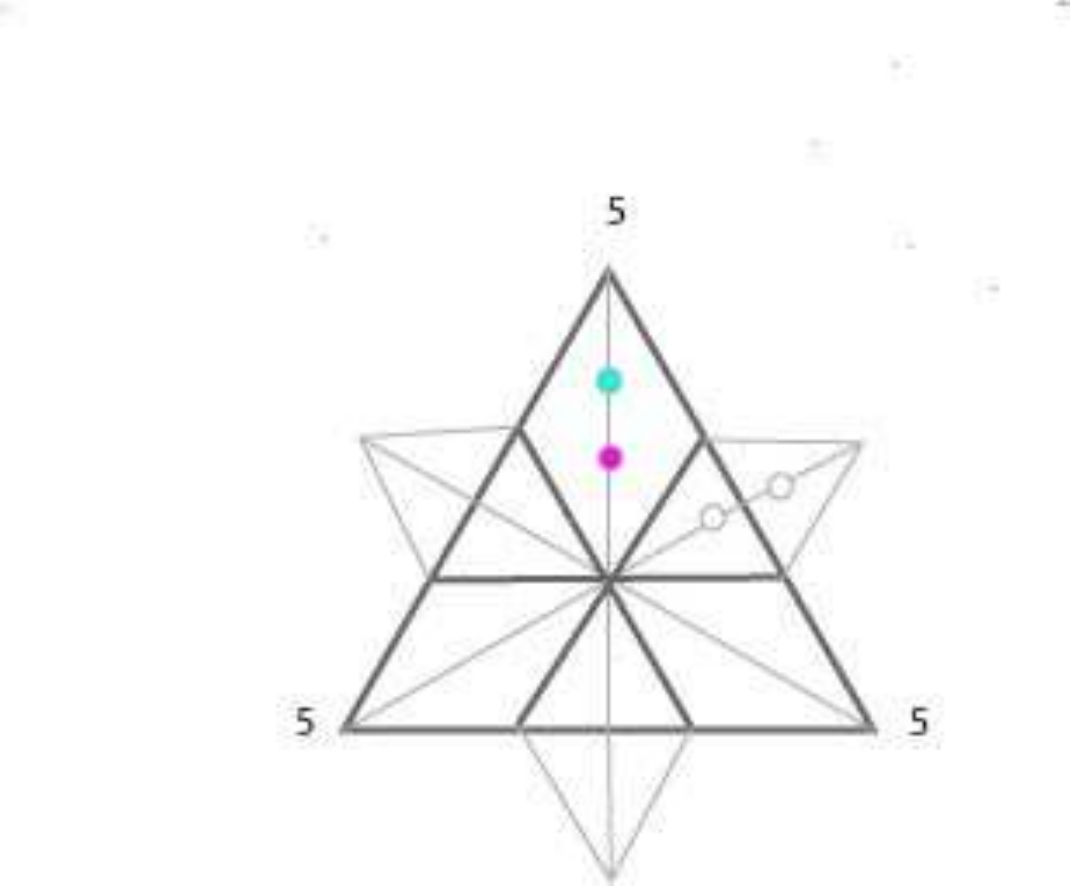}
\end{center}
\caption{\em{Symmetry-corrected MS2 capsids of Type 2 (a)  along a 5-fold axis, (b)  along a 3-fold axis, (c) corresponding tiling decorations.}}
\label{MS22}
\end{figure}

For MS2, there are two inverted candidates in the neighbourhood of a C chain protein, 
one corresponding to the nearest, and the other to the next-to-nearest neighbour. 
\begin{table}[h]
\centering
\begin{tabular}{|c | c | c|}
\hline
chain & colour & deviation from inversion symmetry \\ 
& & (in Angstr\o m) \\ \hline
A  & purple & $5.77$\\
B & cyan & $5.48$ \\
C & white & $5.00$ \\ \hline
 \end{tabular}
\caption{\em The deviation from inversion symmetry (in Angstr\o m) for MS2.}
\label{MS2Table}
\end{table}
The first corresponds to the deviations quoted in Table~\ref{MS2Table} while the second corresponds to a deviation of 5.92 $\AA$ for the C chain. We choose to consider this second option as well because it is not much larger than the maximal deviation for this virus, which is $5.77$ (see the entry for the A chain in Table~\ref{MS2Table}). 
Fig.~\ref{MS21} and   Fig.~\ref{MS22} illustrate capsids with the two types of inversion symmetry corrections, and the corresponding rhombic tilings which will be used in Section 4 to calculate vibrational patterns. Note that the tiling in Fig.~\ref{MS22}(c) coincides with the ideal tiling in Fig.~\ref{fig:icosaidealrhombictiling}.

The case of $T=3$ CK type Tomato Bushy Stunt Virus is straightforward: the analysis of data, based on the file 2TBV.pdb, shows that the capsid is quite further away from a centre of inversion situation, as can be deduced from the deviations presented  in Table~\ref{TBSVtable}. These were calculated by averaging the position of each protein with that of the inverted one (of the same type) closest to it. The resolution at which  TBSV is determined is $2.9\AA$. Table~\ref{TBSVtable} shows that the deviations for TBSV are much larger than those for MS2: whilst the deviation is below $6\AA$  for MS2, it is about $10\AA$  for TBSV.   
Recall from Section 2 that the $T=3$ CK ideal tiling of Fig.~\ref{fig:icosaidealcktiling} does usually not exhibit a centre of inversion either, because the proteins constituting the hexamers differ. 

\begin{figure}[ht]
\begin{center}
\includegraphics[width=6.0cm,keepaspectratio]{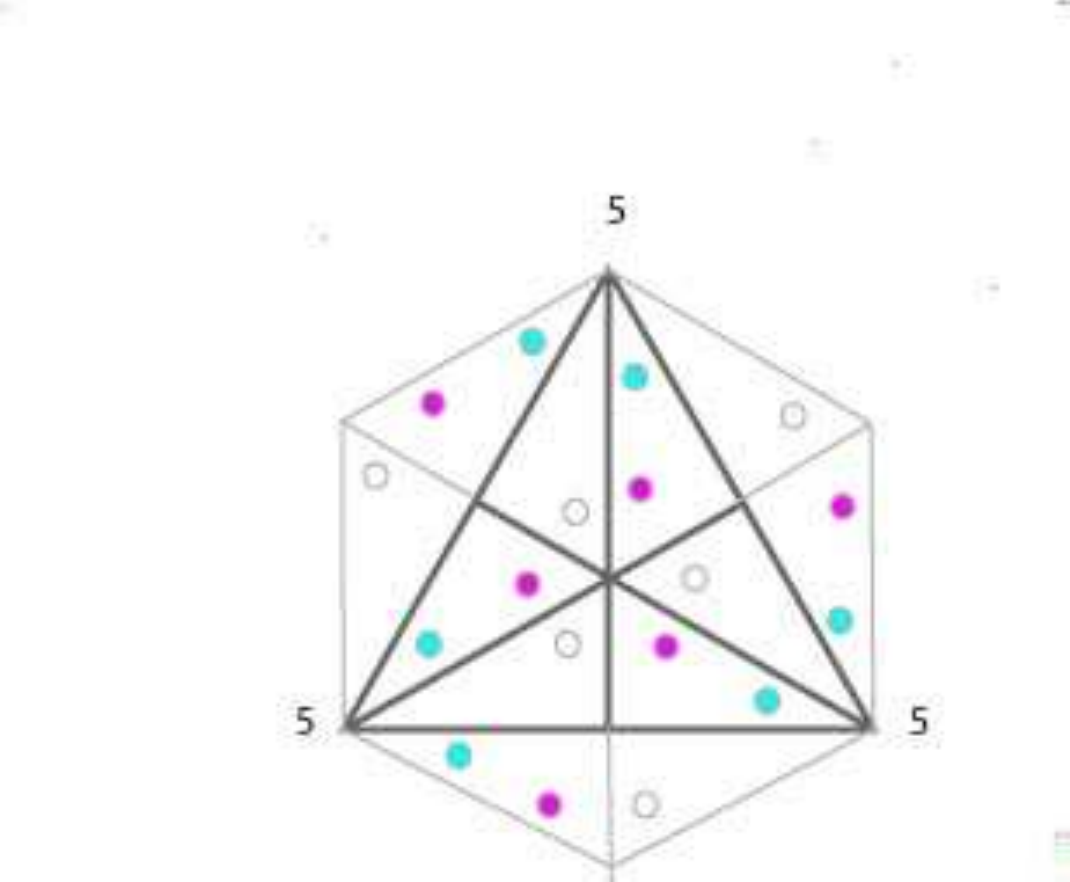}
\end{center}
\caption{\em{The triangular tiling and its decorations for TBSV at the light of experimental data.}}
\label{tilingTBSV}
\end{figure}

We also provide in Fig.~\ref{tilingTBSV} the tiling induced from the experimental data, where the dots represent the centres of mass of individual proteins.
\begin{table}[h]
\centering
\begin{tabular}{|c | c | c|}
\hline
chain & colour & deviation from inversion symmetry \\ 
& & (in Angstr\o m) \\ \hline
A  & purple & $9.66$\\
B & cyan & $10.03$ \\
C & white & $8.86$ \\ \hline
 \end{tabular}
\caption{\em The deviation from inversion symmetry (in Angstr\o m) for TBSV.}
\label{TBSVtable}
\end{table}

\subsection{Simian Virus 40 and Bacteriophage HK97} 
 
\begin{figure}[ht]
\begin{center}
(a)\includegraphics[width=5cm,keepaspectratio]{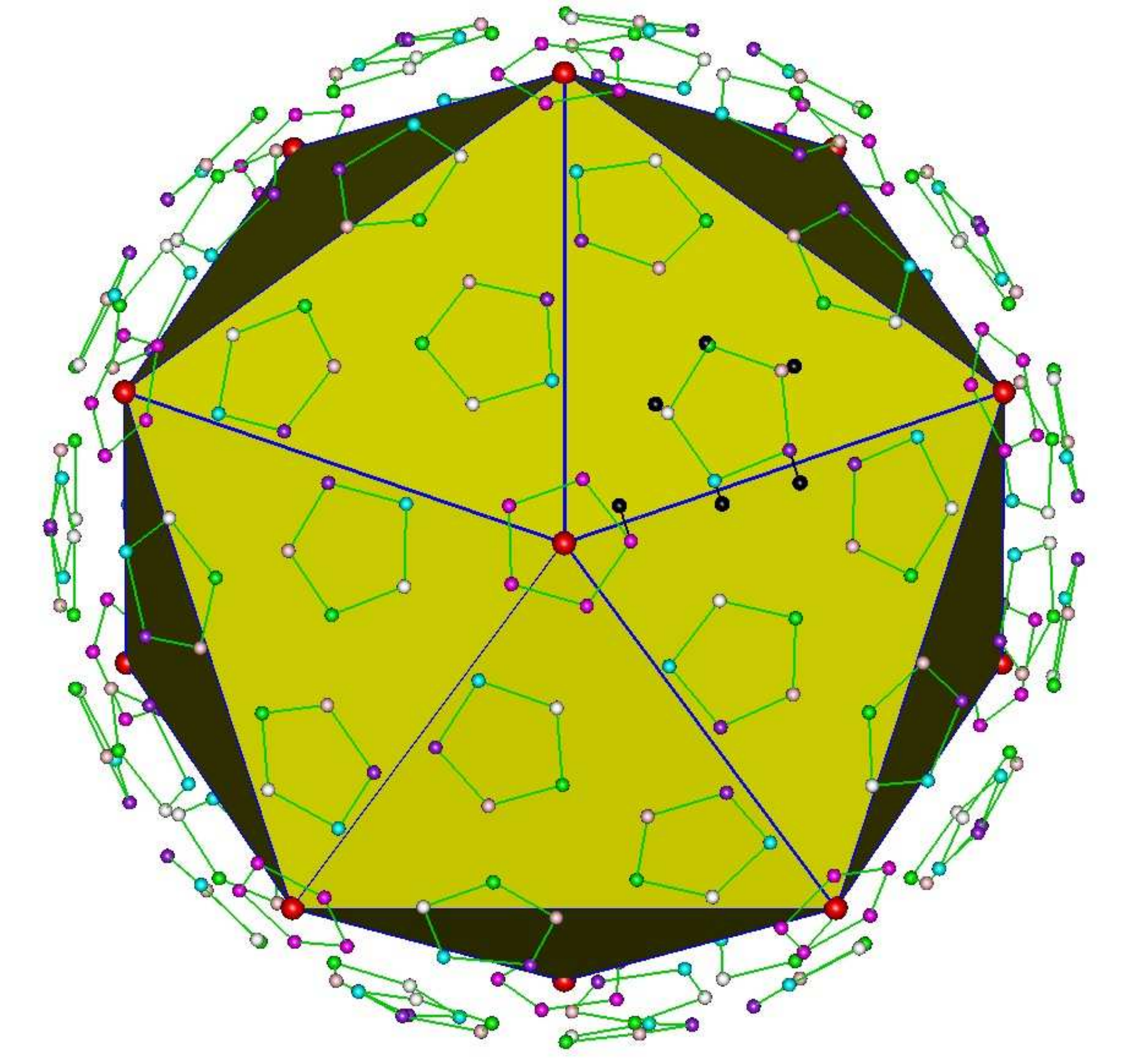}
(b)\includegraphics[width=4.7cm,keepaspectratio]{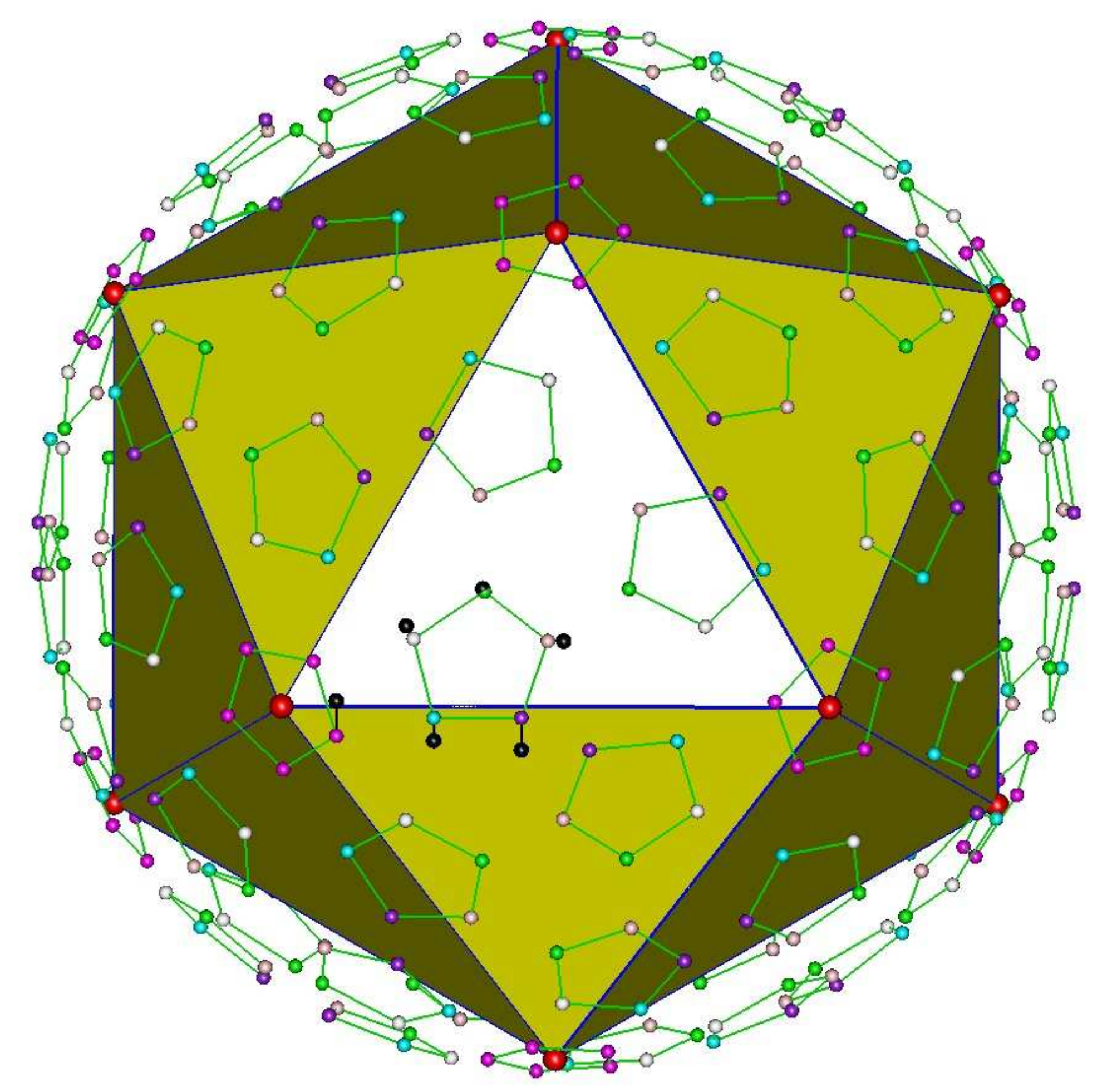}
(c)\includegraphics[width=4.2cm,keepaspectratio]{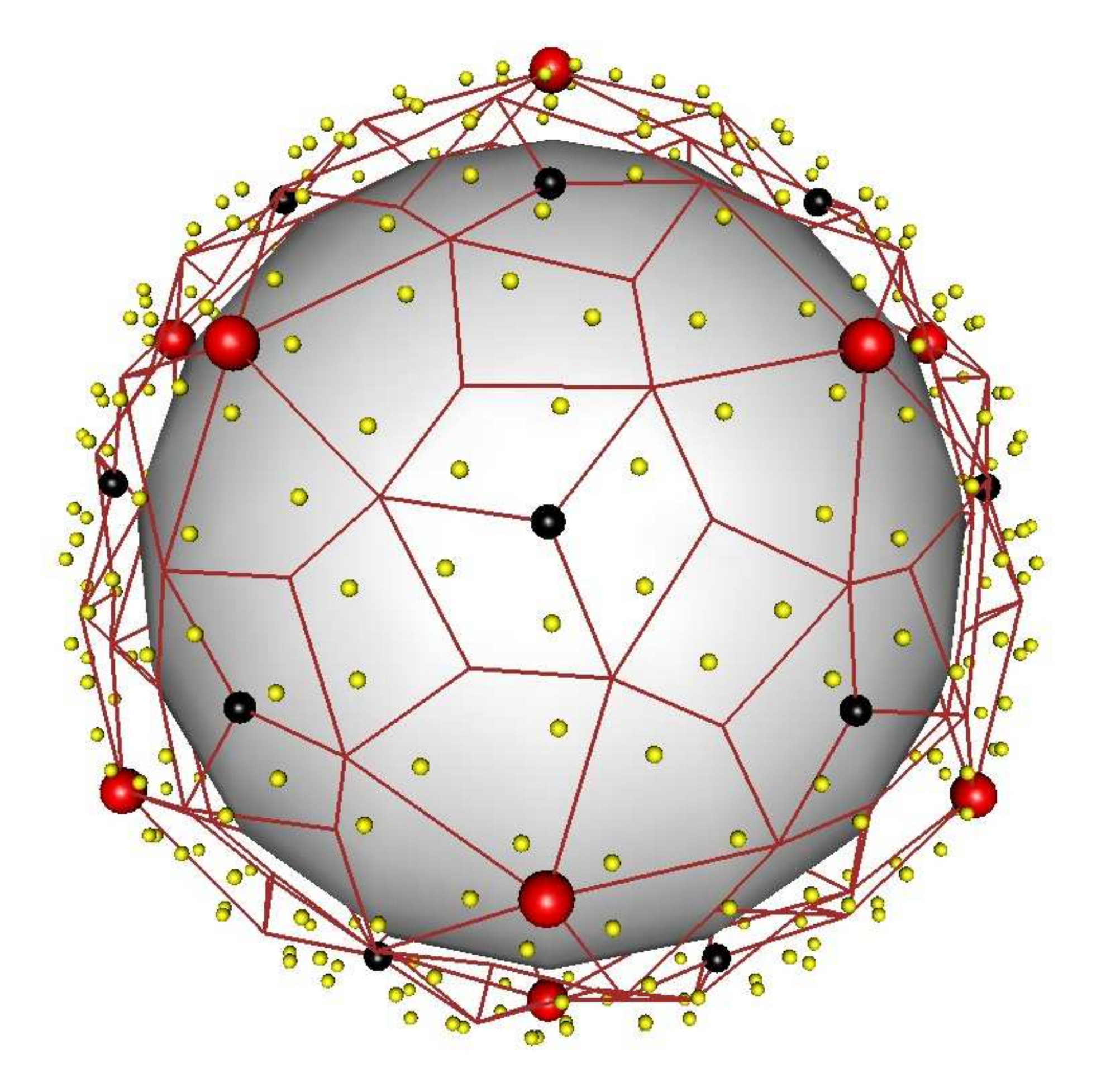}
\end{center}
\caption{\em{The calculated centres of mass for the capsid proteins in chains A, B, C, D, E and F of Simian Virus 40 are represented by big dots in purple, cyan, white, green, pink and magenta respectively,  while the small black dots correspond to the distribution of centres of mass after inversion. Views from above a 5-fold (a) and a 3-fold (b) symmetry axes are provided; (c) Rhomb and kite tiling shown in comparison with experimental data. Red dots represent axes of five-fold and black dots axes of three-fold icosahedral symmetry.}}
\label{SVA}
\end{figure}

 We now turn our attention to SV40, a $T=7d$ capsid with 360 capsid proteins organised in first approximation according to the rhomb and kite tiling presented in Fig.~\ref{fig:sv40bis}. This is an all-pentamer capsid consisting of six chains. Based on the 1SVA.vdb file, we represent in Fig.~\ref{SVA} the centres of mass of the A, B, C, D, E and F chains in purple, cyan, white, green, pink and magenta respectively.  

We use this information to derive the qualitative locations of these vertices with respect to a  $T=7d$ rhomb and kite tiling. The result is shown  in Fig.~\ref{SVA}(c). In Fig.~\ref{SVA}(a) and (b), we show in black some of the vertices obtained by inversion from vertices located on the far side of the capsid. The distances between these black vertices and the vertices to which they are connected by a black line correspond to twice the deviation from inversion symmetry for these vertices. 

The deviation from inversion symmetry can be determined for each chain, and we obtain the figures given in Table~\ref{SVATable}. Although the resolution at which  SV40 is determined is 3.1 $\AA$, we argue that the small deviations in Table~\ref{SVATable} are consistent with the assumption that the viral capsid has an approximate centre of inversion. 
\begin{table}[h]
\centering
\begin{tabular}{|c | c | c|}
\hline
chain & colour & deviation from inversion symmetry \\ 
& & (in Angstr\o m) \\ \hline
A  & purple & $7.07$\\
B & cyan & $4.99$ \\
C & white & $3.06$ \\ 
D  & green & $0.70$\\
E & pink & $3.25$ \\
F & magenta & $7.64$ \\ \hline 
 \end{tabular}
\caption{\em The deviation from inversion symmetry (in Angstr\o m) for SV40.}
\label{SVATable}
\end{table}

\begin{figure}[ht]
\begin{center}
(a)\includegraphics[width=5.2cm,keepaspectratio]{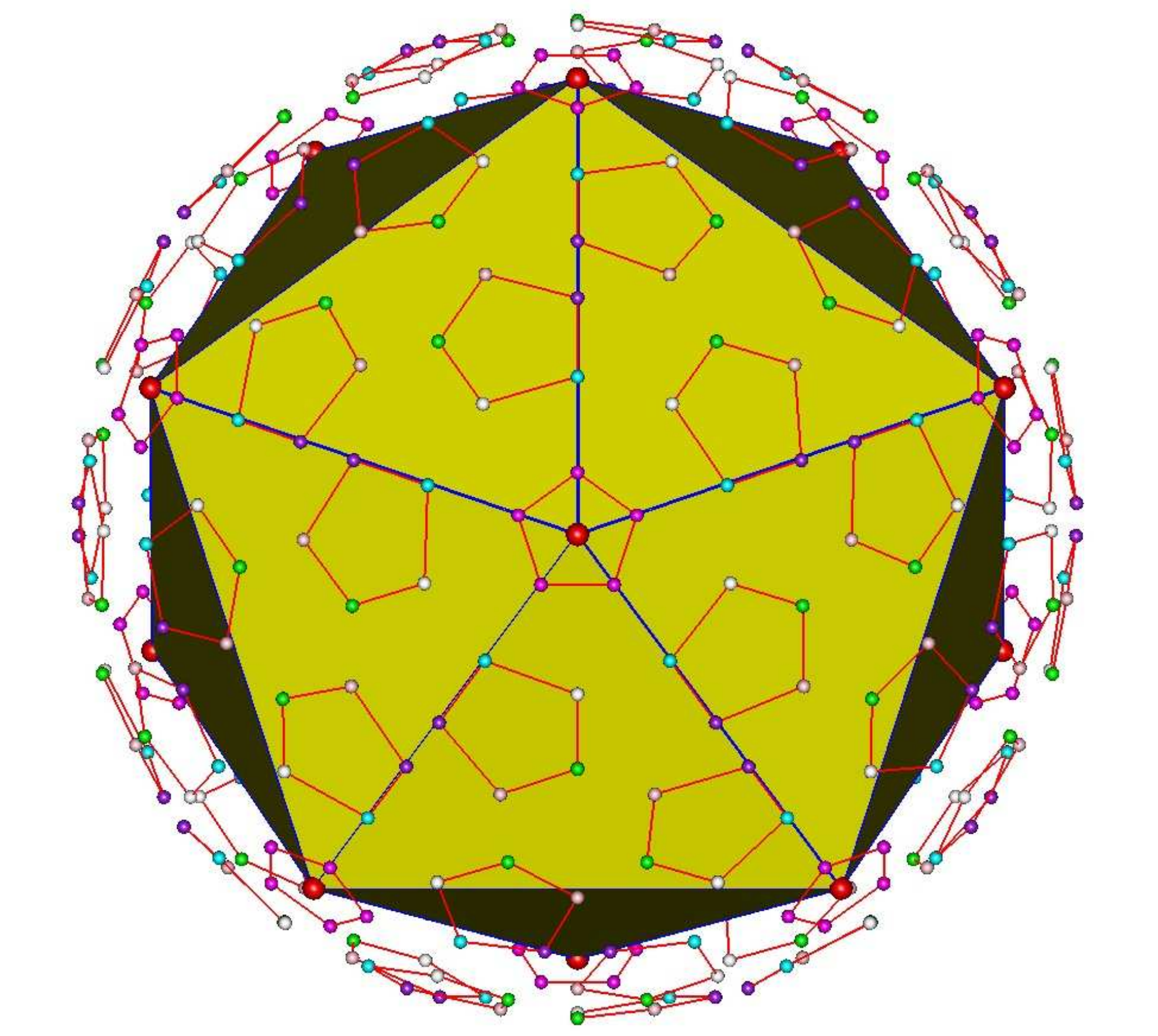}
(b)\includegraphics[width=4.7cm,keepaspectratio]{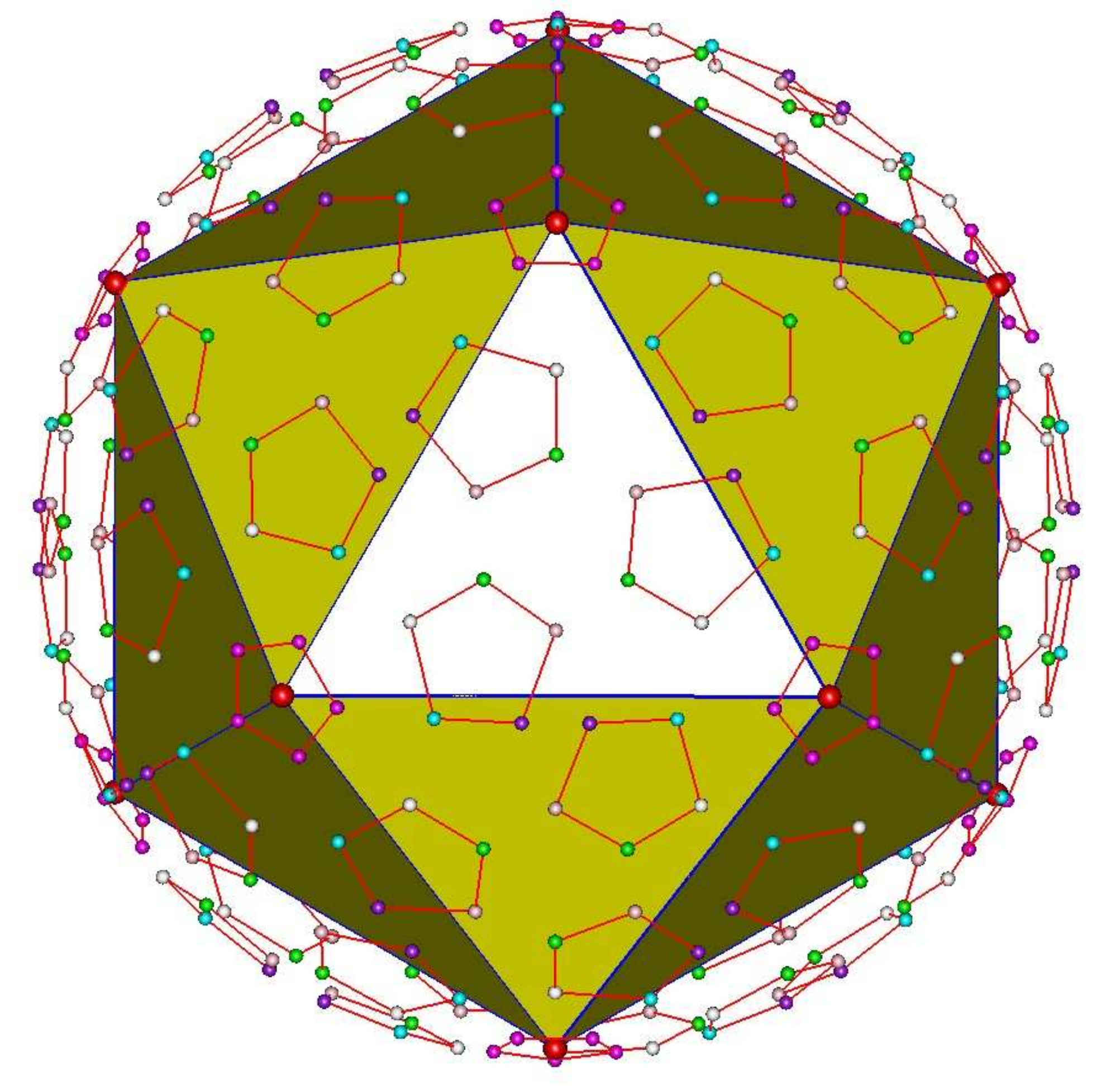}
(c)\includegraphics[width=4.0cm,keepaspectratio]{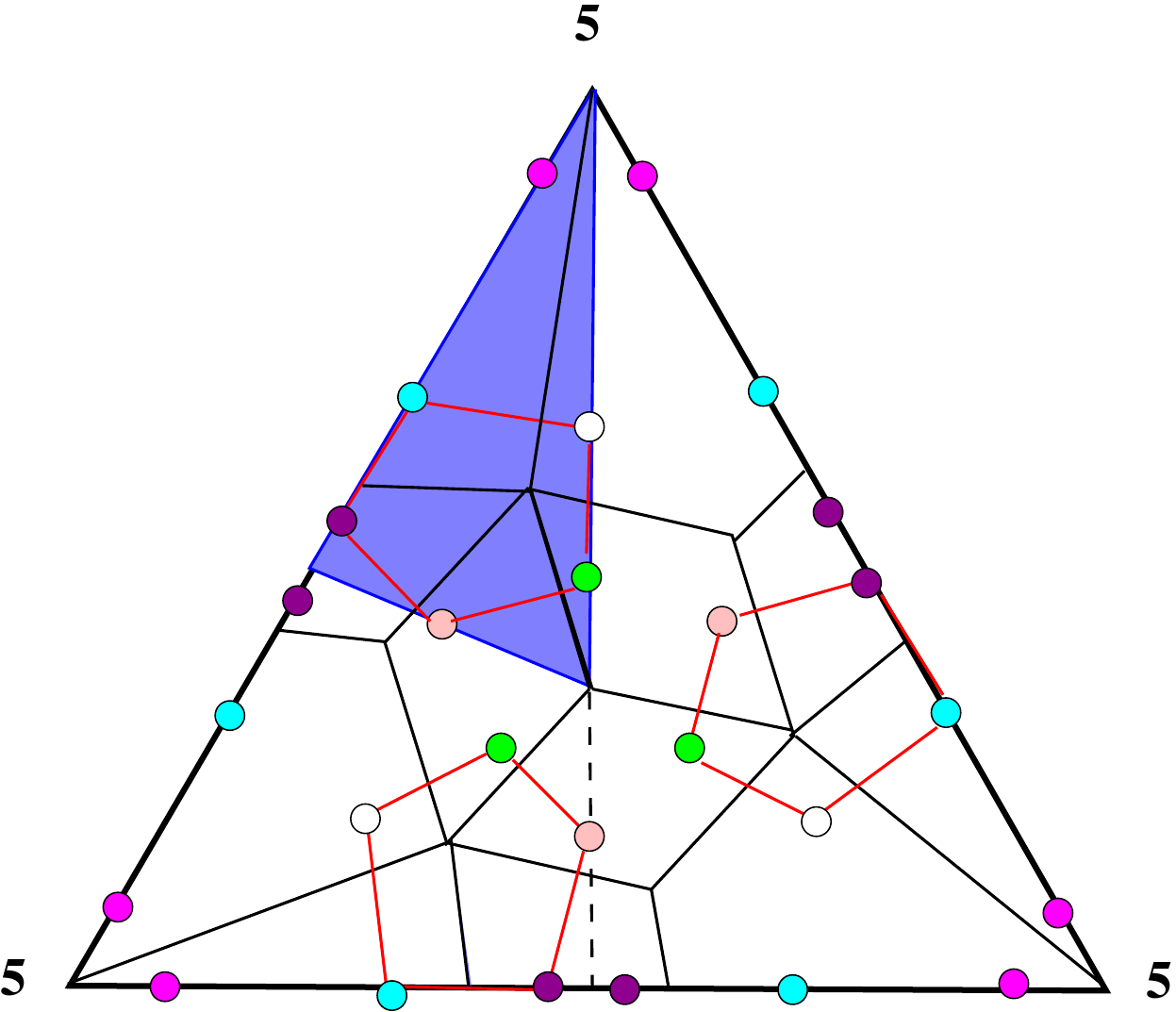}
\end{center}
\caption{\em{The symmetry corrected versions for SV40: 3d representation of the location of proteins centres of mass in a five-fold (a) and  three-fold (b) view; (c) the corresponding rhomb and kite tiling and its decorations.}}
\label{SVA1}
\end{figure}

As in the case of Bacteriophage MS2, we calculate the symmetry-corrected version of the tiling by averaging the position of each protein with that of the inverted one (of the same chain) closest to it. The results are presented in Fig.~\ref{SVA1}.

We end with Bacteriophage HK97, whose $T=7\ell$ capsid has 420 proteins belonging to 7 different chains, organised in pentamers around the global 5-fold symmetry axes of the underlying icosahedron and hexamers everywhere else. Using the data from the file 2fte.vdb, we have again calculated the centres of mass of all capsid proteins involved and compared their distribution with the distribution obtained by inversion. Table~\ref{FTETable} lists the deviations of each chain from a distribution with inversion symmetry, and, given the relatively large numbers,  one cannot reasonably argue that the HK97 capsid has a centre of inversion, even approximately.

\begin{table}[h]
\centering
\begin{tabular}{|c | c | c|}
\hline
chain & colour & deviation from inversion symmetry \\ 
& & (in Angstr\o m) \\ \hline
A  & purple & $16.64$\\
B & cyan & $2.33$ \\
C & white & $10.92$ \\ 
D  & green & $17.84$\\
E & pink & $13.3$ \\
F & magenta & $0.61$ \\ 
G&blue&0.61\\\hline 
 \end{tabular}
\caption{\em The deviation from inversion symmetry (in Angstr\o m) for HK97.}
\label{FTETable}
\end{table}

\setcounter{equation}{0}
\section{Group theoretical properties of Raman modes}
Our aim is to provide qualitative information on the vibrational modes of the viral capsids described in the previous section. As explained there, we work within a  coarse-grain approximation which consists in substituting each capsid protein by a point mass. This point mass is located either at the centre of mass of the given protein, or at the midpoint between its centre of mass and that of an inverted protein of the same chain in its immediate neighbourhood. The former approximation is chosen when the capsid protein distribution deviates significantly from a distribution invariant under inversion (TBSV and HK97), while the latter is adopted for MS2 and SV40 as their capsid protein distributions are inversion invariant to good approximation. This particular coarse-graining reduces considerably the number of degrees of freedom which must be treated in the vibration analysis. Given a virus with $N$ capsid proteins, the system depends on $3N$ degrees of freedom. Strictly speaking, our capsids are distributions of point masses calculated according to one of the two  prescriptions above, but we will, in the remainder of this paper, speak of distributions of capsid proteins although the word `protein' here must be taken in a loose sense.

Precise numerical values for the frequencies of normal modes of vibrations can only be obtained by diagonalisation of the force matrix, which encodes the interactions between the point particles of our approximation. Group theory provides an elegant route to the diagonalisation \cite{Peeters Taormina}, and together with Viral Tiling ideas, offers valuable insights in the description of patterns of normal modes, as we proceed to show.

We pay particular attention to the normal modes of vibration which can be detected via  Raman spectroscopy \cite{Raman}, as this technique is, unlike infrared spectroscopy, well-suited for measuring frequency spectra in aqueous samples such as living organisms.

The following group theoretical considerations are standard and can be found, for instance, in \cite{Cornwell}.
\subsection{Displacement representation of $H_3$ or ${\cal I}$ for generic capsids}

The first step in this analysis involves the decomposition of the displacement representation of the capsid into a direct sum of irreducible representations of the icosahedral group $H_3$ or its proper rotation subgroup ${\cal I}$.

The former is used whenever the capsid has an approximate centre of inversion, and the latter when it does not.

The displacement representations of the icosahedral group $H_3$ (resp. of its subgroup ${\cal I}$) for  viruses or phages with N capsid proteins  consist of 120 (resp. 60) matrices $\Gamma^{displ}_{3N}(g),\,g \in H_3 \,(\rm resp.\,{\cal I}$) of size $3N \times 3N$, which encode how proteins are interchanged under the action of each element $g$, as well as how the displacements of each protein from the equilibrium position are rotated under the action of $g$.
 The latter information is gathered in $3 \times 3$ rotation matrices $R(g)$ which form an irreducible  representation of $H_3$ (resp. ${\cal I})$, while the former is encoded in permutation matrices $P(g)$ of size $N \times N$, so that we have
\be
\Gamma^{displ}_{3N}(g)=P(g) \otimes R(g), \qquad \qquad g \in H_3 \,({\rm resp.}\, {\cal I}).
\ee
The permutation matrices $P(g)$ act on vectors whose components are the vector positions $\vec{r}^{\,\,e}_i, i=1,..,N$ of the N proteins at equilibrium. The entry  $P_{ij}(g)$ of the permutation matrix is 1 if 
$\vec{r}^{\,\,e}_j$ is mapped on $\vec{r}^{\,\,e}_i$ by $g$, and is zero otherwise.

\subsection{Decomposition of $\Gamma^{displ}$ into irreducible representations}

There exists a matrix $U$ - whose detailed form is not needed for our present analysis - which transforms the displacement representation into a finite sum of irreducible representations $\Gamma^p$ of $H_3$ (resp. $ {\cal I}$):
\be
U\Gamma^{displ}_{3N}(g)U^{-1}=\Gamma^{displ\,\,'}_{3N}(g)=\oplus_p n_p\Gamma^p(g),
\ee
where the multiplicities $n_p$ are obtained via the following character formula
\be \label{multiplicities}
n_p=\frac{1}{{\rm dim\, H_3}}\sum_{g \in H_3} \chi^{displ}(g)^*\,\chi^p(g)\qquad {\rm or}\qquad n_p=\frac{1}{{\rm dim\, {\cal I}}}\sum_{g \in  {\cal I} } \chi^{displ}(g)^*\,\chi^p(g).
\ee
The characters $\chi^p(g)$ of irreducible representations of the icosahedral group are listed in Table~\ref{charTable}, while the characters of the displacement representations $\chi^{displ}(g)$ are obtained by inspection of the displacement representation considered. Note that, in view of the very definition of the permutation matrices $P(g)$ given in the previous subsection, and the fact that the characters of a representation are the traces of its constituent matrices, one has 
\be \label{traces}
\chi^{displ}(g)= {\rm Tr}\,(P(g))\,{\rm Tr}\,(R(g))=\pm ({\rm number\, of\, proteins\, unmoved\, by\,}g)\cdot (1+2\cos\theta),
\ee
where $\theta$ is the angle of the proper rotation associated with $g$, and the minus sign is taken when $g \in H_3 \setminus {\cal I}$. So $\chi^{displ}(g)$ is zero when $\theta=\frac{2\pi}{3}$ or whenever $g$ is such that no protein of a given capsid  is kept fixed under its action. 

The decomposition of the displacement representation of  a given capsid boils down to the knowledge of the coefficients $n_p$ in \eqref{multiplicities} which, in view of the expression \eqref{traces}, are non zero whenever at least one capsid protein is unmoved under the action of an element $g$ (and $\theta \neq \frac{2\pi}{3}$). 

Before dwelling into the particulars of the four capsids we have chosen to analyse here, let us enumerate a few properties of generic icosahedral capsids relevant to the calculation of the decomposition coefficients $n_p$.

We set the origin of coordinates at the centre of the icosahedron which supports the distribution of capsid proteins, so that all symmetry axes go through $(0,0,0)$.

{\bf Property 1}: Consider a virus shell with N capsid proteins distributed according to icosahedral symmetry. Then the identity element $e$ of the icosahedral group keeps all N proteins trivially unmoved and  Tr $P(e)=N$. The corresponding  rotation $R(e)$ has trace $+3$ and $\chi^{displ}(e)=3N$.

{\bf Property 2}: A capsid protein is unmoved by a proper 5-fold, 3-fold or 2-fold rotation if  and only if it is located on the 5-, 3- or 2-fold symmetry axis respectively.

{\bf Property 3}: Consider a viral capsid exhibiting invariance under inversion, and identify an arbitrary 2-fold axis of the capsid. Then any capsid protein lying in the plane orthogonal to that 2-fold axis is unmoved by the element $g_0g_2=g_2g_0$ which consists in a 2-fold rotation about the chosen 2-fold axis followed by an inversion, or vice-versa. 

\begin{figure}[ht]
\begin{center}
(a)\includegraphics[width=5.1cm,keepaspectratio]{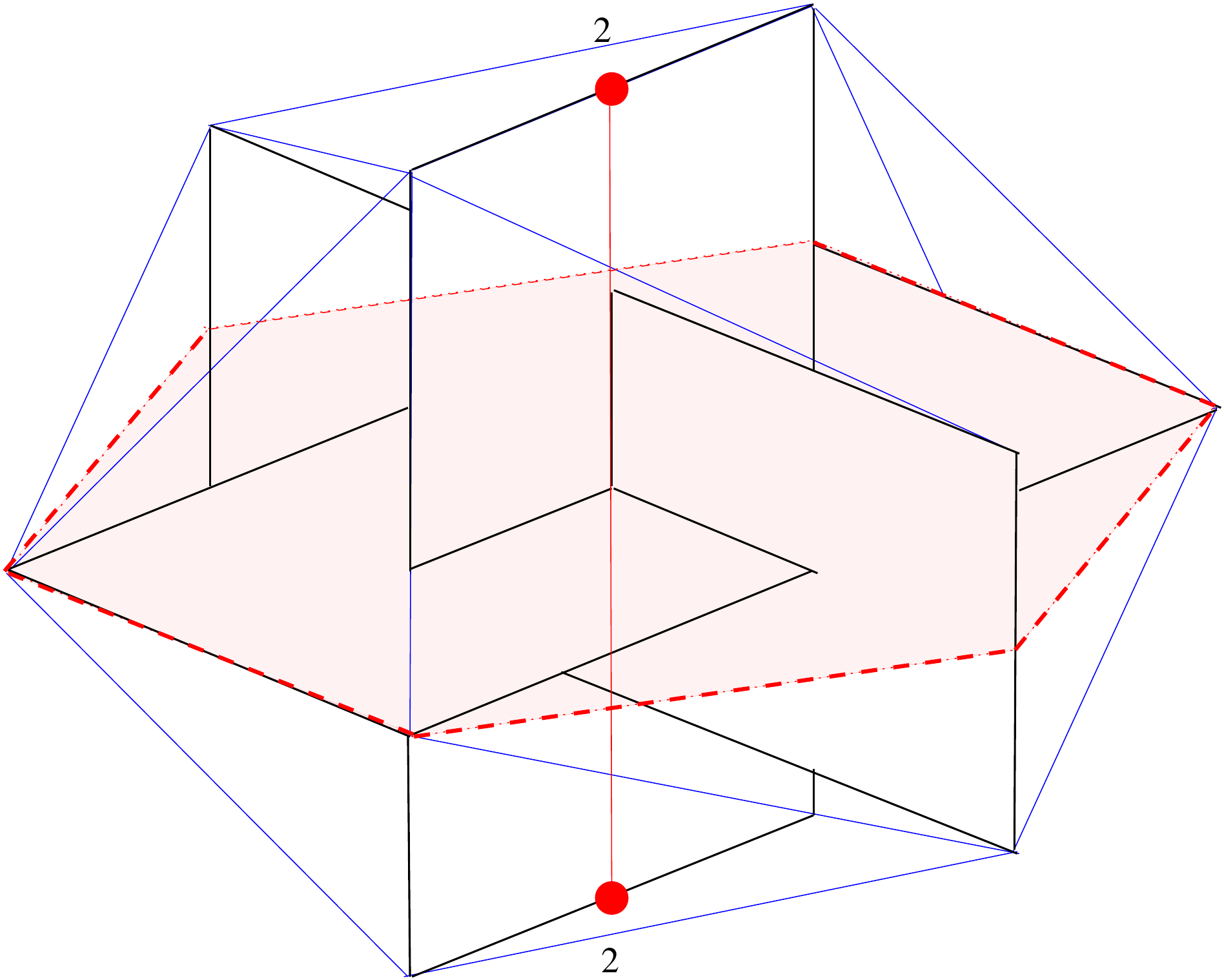}
(b)\includegraphics[width=8.1cm,keepaspectratio]{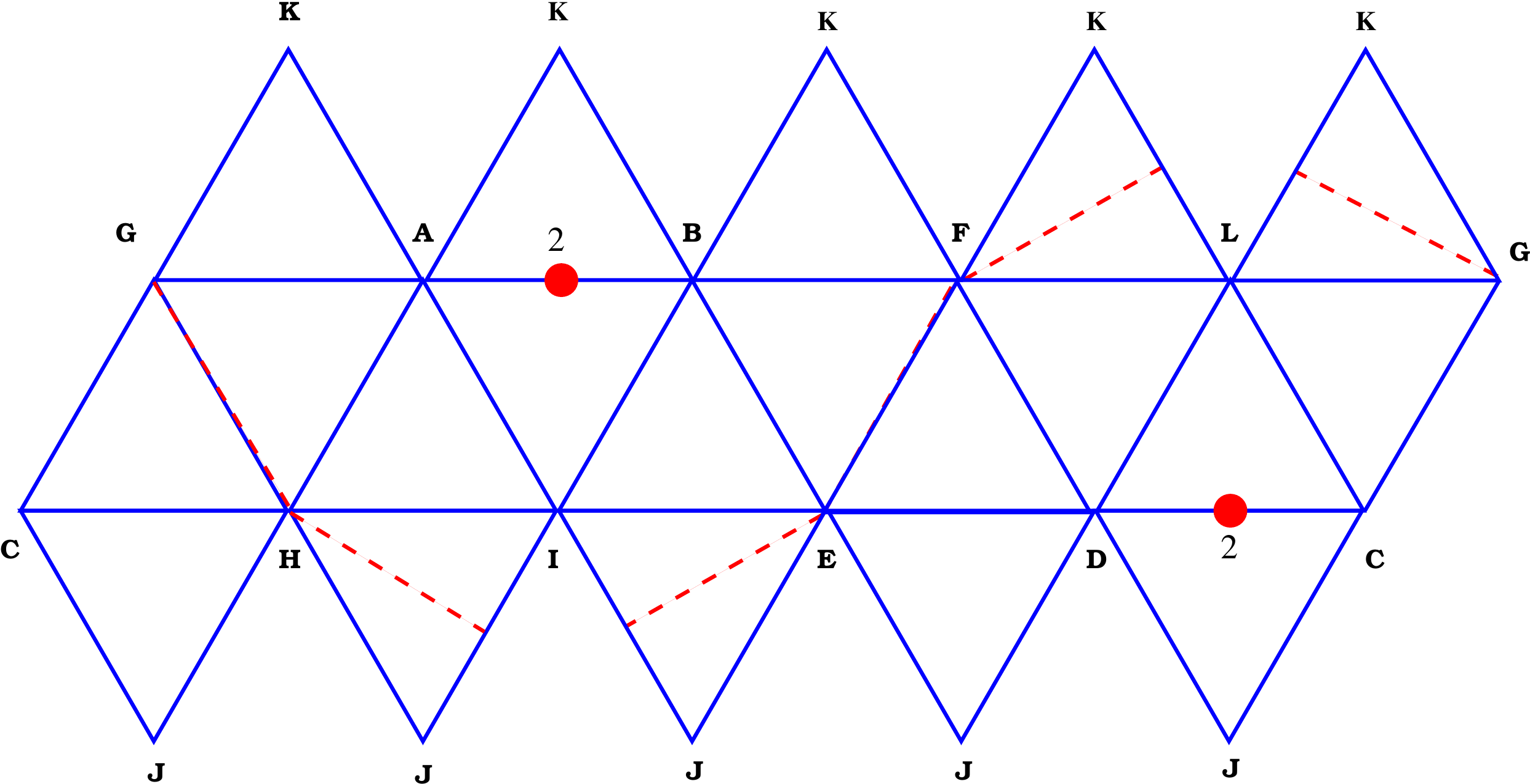}
\end{center}
\caption{\em{(a) An hexagonal portion of the plane orthogonal to a given (red) 2-fold symmetry axis is shaded in red in a representation of the icosahedron involving three orthogonal golden rectangles, with the icosahedron suprimposed in blue, (b) The perimeter of the hexagonal portion projected  on a planar template of the icosahedron. The red dots represent the intersection of the 2-fold axis with the icosahedron.}}
\label{fig:unmoved}
\end{figure}

Fig.~\ref{fig:unmoved} shows the 12 vertices of an icosahedron as the vertices of three golden rectangles orthogonal to each other. A 2-fold axis is highlighted in red and an hexagonal portion of the plane orthogonal to that axis is shaded in red . The perimeter of the hexagon is the intersection of the icosahedron with the orthogonal plane, and proteins unmoved by  $g_0g_2$ must be located on this perimeter.
 \subsection{Raman active modes of the four viral capsids}
 
 Let us now look at the details of the decomposition into irreducible representations of the displacement representation of the four chosen capsids.

 {\bf {\em MS2 type 1:}}
 this capsid has an approximate centre of inversion and therefore, we use the full icosahedral group $H_3$. According to the decorations in Fig.~\ref{MS21} (c):
 \begin{enumerate}
 \item [(i)] No protein is located on a 5-, 3- or 2-fold symmetry axis, hence no proper rotation bar the identity leaves any protein unmoved. The identity $e$ however keeps all 180 proteins trivially unmoved and $\chi(P(e))\equiv \rm {Tr} P(e)=180$. Correspondingly, the rotation $R(e)$ has trace $+3$ and $\chi^{displ}(e)=540$.
 \item [(ii)] All proteins on the plane orthogonal to a given 2-fold symmetry axis are distributed according to Fig~\ref{icosaunmovedMS2}(a). There are 12 proteins on that particular plane, hence $\chi(P(g_0g_2))\equiv \rm{Tr}\, P(g_0g_2)=12$ while $\chi(R(g_0g_2))\equiv \rm{Tr}\, R(g_0g_2)=-1 \times (1+2\cos \pi)=1$, so that $\chi^{displ}(g_0g_2)=  12 \times 1=12$. There are 15 ways to choose a 2-fold axis on the icosahedron, hence 15 elements of $H_3$ are of type $g_0g_2$, which must be accounted for when calculating
 the coefficients $n_p$ from \eqref{multiplicities}.
\begin{figure}[ht]
\begin{center}
(a)\includegraphics[width=7.1cm,keepaspectratio]{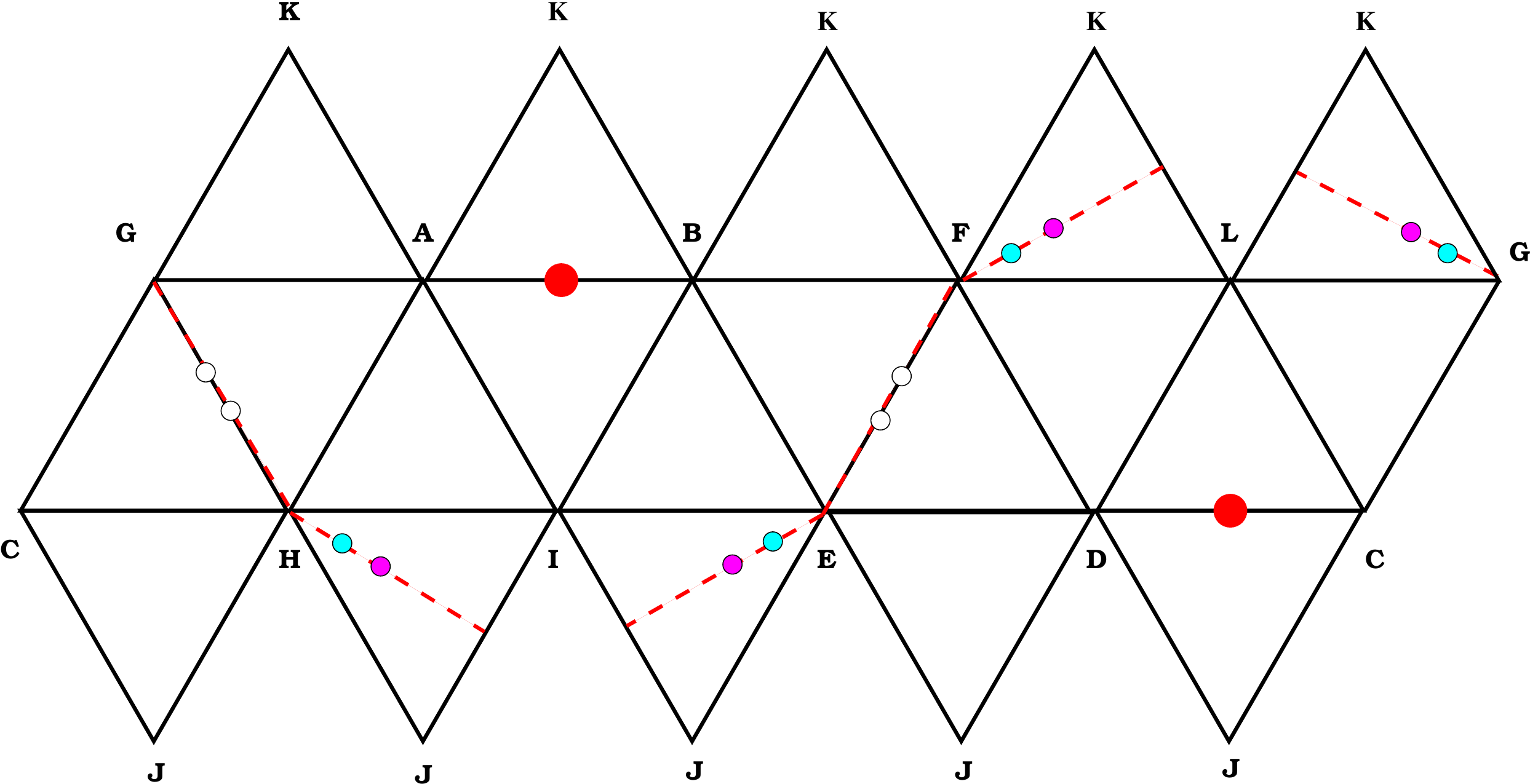}
(b)\includegraphics[width=7.1cm,keepaspectratio]{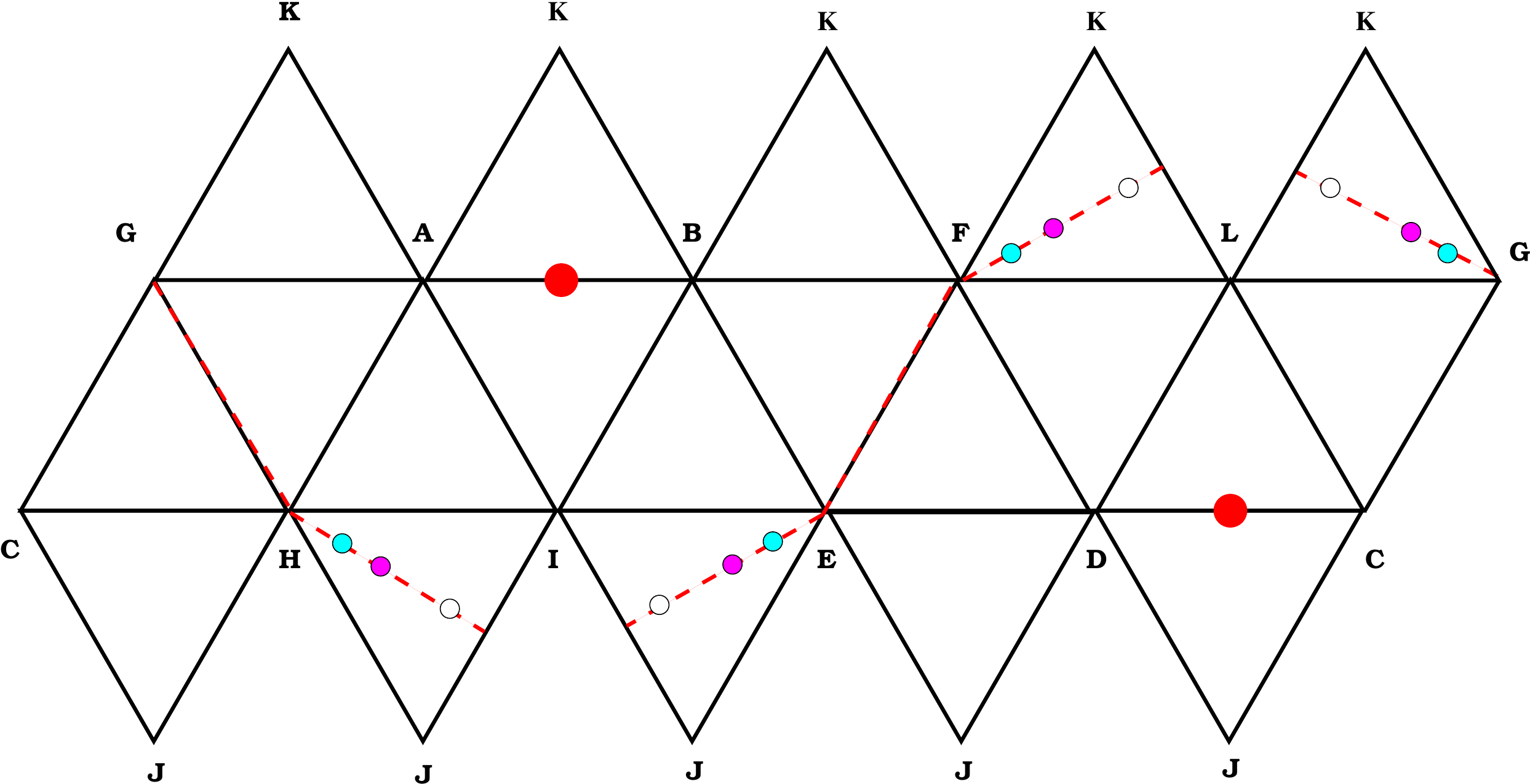}
\end{center}
\caption{\em{(a) The 12 MS2 capsid proteins located on the plane orthogonal to a 2-fold symmetry axis represented by the red dots, when considering the symmetry - corrected version pictured in Fig.~\ref{MS21}(c); (b) The 12 MS2 capsid proteins located on the plane orthogonal to a 2-fold symmetry axis represented by the red dots, when considering the symmetry - corrected version pictured in Fig.~\ref{MS22}(c)}.}
\label{icosaunmovedMS2}
\end{figure}
\item [(iii)] No other element of $H_3$ leaves any capsid protein of Bacteriophage MS2 type 1 unmoved, so that the coefficients are given by
\be
n_p^{MS21}=\frac{1}{120}\left\{540 \,\chi^p(e)\,+\,180\,\chi^p(g_0g_2) \right\}, \quad p=1_{\pm}, 3_{\pm}, 3'_{\pm},  4_{\pm}, 5_{\pm}.
\ee
Using the information in Table~\ref{charTable}, one obtains the following decomposition of the displacement representation into irreducible representations of $H_3$, labelled
$\Gamma_{\pm}^1, \Gamma_{\pm}^3, \Gamma_{\pm}^{3'}$, $\Gamma_{\pm}^4, \Gamma_{\pm}^5$, 
\be \label{decompMS21}
\Gamma^{displ\,'}_{540,MS21}= 6\Gamma^1_+ + 12\Gamma^3_+ + 12\Gamma^{3'}_+ +18\Gamma^4_+ +
24\Gamma^5_+ 
+ 3\Gamma^1_- + 15\Gamma^3_- + 15\Gamma^{3'}_- + 18\Gamma^4_- +
21\Gamma^5_-.
\ee
The above describes all modes of vibration of the $T=3$ icosahedral capsid modelling Bacteriophage MS2 type 1, including rotations and translations of the capsid as a whole. These six degrees of freedom belong to the $\Gamma^3_+$ and $\Gamma^3_-$ irreducible representations of $H_3$ respectively, and must be subtracted from \eqref{decompMS21} in order to classify the normal modes of vibration of the capsid.
We thus have
\be
\Gamma_{540,MS21}^{vib}= 6\Gamma^1_+ + 11\Gamma^3_+ + 12\Gamma^{3'}_+ +18\Gamma^4_+ +
24\Gamma^5_+ 
+ 3\Gamma^1_- + 14\Gamma^3_- + 15\Gamma^{3'}_- + 18\Gamma^4_- +
21\Gamma^5_-.
\ee
It remains to identify, among the above modes, those which are potentially Raman active. Recall that Raman spectroscopy involves placing a molecule (here a virus), which vibrates at frequencies $\nu_{internal}$, in a time-varying electric field produced, for instance, by a monochromatic laser of frequency $\nu_{laser}$. One then looks for frequency shifts $\nu_{laser} \pm \nu_{internal}$ in the light which scatters inelastically from the molecule. Such shifts only occur if the internal motion of the molecule induces a change in its polarizability, which is a 2-rank symmetric tensor  $\alpha_{ij}, i,j =1,2,3$ entering in the definition of the dipole moment $\mu_i$ of the molecule induced by the presence of an electric field $E_i$ ($\mu_i=\alpha_{ij}E_j$) \cite{Herzberg}. Given that the components of the polarizability tensor transform as \footnote{ $x_i, i=1,2,3$ are the coordinates of a point in 3-space.} $x_i^2\,(i=1,2,3)$ and $x_i\,x_j$, which belong to the $\Gamma^1_+$ and the $\Gamma^5_+$ irreducible representations of $H_3$ (see Table~\ref{charTable}), we have
\be
\Gamma_{540,MS21}^{Raman}=6\Gamma^1_+ + 24\Gamma^5_+ ,
\ee
i.e. the Raman active fundamental levels are the six non-degenerate levels belonging
to $\Gamma^1_+$ and the twenty-four five-degenerate levels belonging to $\Gamma^5_+$. The infrared active modes belong to $\Gamma_+^3$ and $\Gamma_-^3$, and hence are completely decoupled from the Raman active modes, as expected from a capsid with a centre of inversion.
 \end{enumerate}\vskip 1cm
{\bf  {\em MS2 type 2:}} The analysis leads to the same decomposition pattern as the MS2 type 1. According to the decorations in Fig.~\ref{MS22} (c), the number of proteins left unmoved by elements of $H_3$ of type $g_0g_2$ for a given 2-fold rotation is 12 again, although their distribution on the orthogonal plane is different from the previous case.
 This does not affect the qualitative analysis of Raman modes, but may be of importance when calculating the frequencies of the normal modes of vibration \cite{Peeters Taormina}. We thus have,
 \be
 \Gamma_{540,MS22}^{vib}=\Gamma_{540,MS21}^{vib}
 \ee
 and
 \be 
 \Gamma_{540,MS22}^{Raman}=\Gamma_{540,MS21}^{Raman}.
 \ee
 \vskip 1cm
{\bf  {\em TBSV:}} This viral capsid does not exhibit  a centre of inversion, even approximate, and the relevant group for our analysis is  the subgroup ${\cal I}$ of the full icosahedral group. The only element which leaves proteins unmoved is the identity, under which all 180 proteins are unmoved. The coefficients of the decomposition of the displacement representation in this case are
 \be
n_p^{TBSV}=\frac{1}{60}\left\{540 \,\chi^p(e)) \right\}, \quad p=1_{+}, 3_{+}, 3'_{+},  4_{+}, 5_{+}.
\ee
 We therefore have the decomposition \footnote{ This is a special case of the formula giving the decomposition of the $3N$-dimensional displacement representation of a viral capsid with $N$ capsid proteins into irreducible representations of the group of proper rotations ${\cal I}$, namely $$ \Gamma^{displ\,'}_{3N}=\frac{3N}{60}\left\{  \Gamma^1_+ + 3\Gamma^3_+ + 3\Gamma^{3'}_+ +4\Gamma^4_+ +
5\Gamma^5_+ 
\right\}.$$}
 \be
 \Gamma^{displ\,'}_{540,TBSV}= 9\Gamma^1_+ + 27\Gamma^3_+ + 27\Gamma^{3'}_+ +36\Gamma^4_+ +
45\Gamma^5_+ ,
\ee
and once the rotations and translations are subtracted, we obtain
 \be
 \Gamma_{540,TBSV}^{vib}= 6\Gamma^1_+ + 25\Gamma^3_+ + 27\Gamma^{3'}_+ +36\Gamma^4_+ +
45\Gamma^5_+. 
\ee
The Raman modes of vibrations are
\be 
 \Gamma_{540,TBSV}^{Raman}= 6\Gamma^1_+ + 45\Gamma^5_+ ,
 \ee
 i.e. the Raman active fundamental levels are the six non-degenerate levels belonging
to $\Gamma^1_+$ and the forty-five five-degenerate levels belonging to $\Gamma^5_+$. These modes are not infrared active. Note that the Raman signatures of Bacteriophage
MS2 and Tomato Bushy Stunt Virus are different, due to the existence of an approximate centre of inversion for MS2, but not for TBSV.
\vskip 1cm

\begin{figure}[ht]
\begin{center}
\includegraphics[width=8.1cm,keepaspectratio]{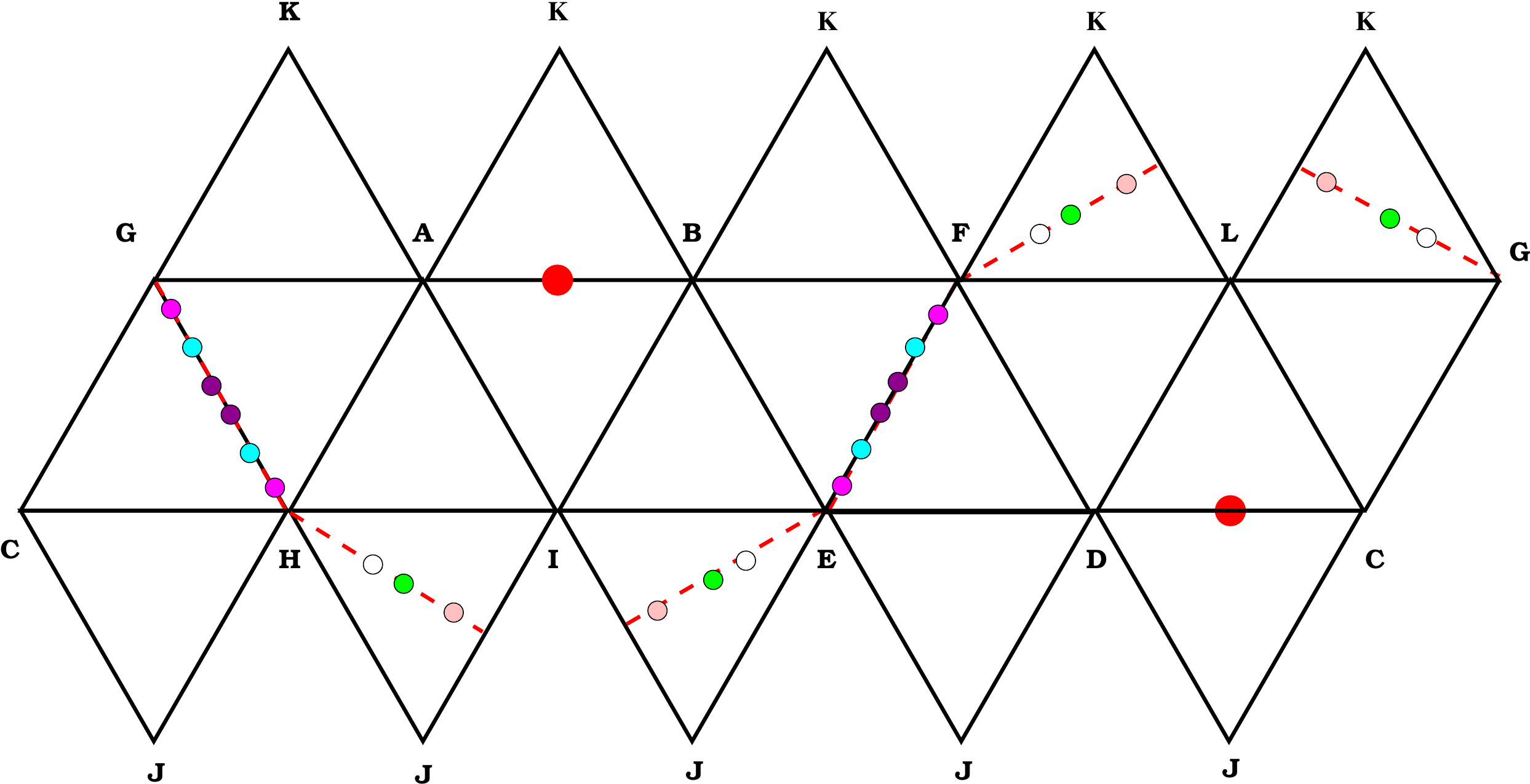}
\end{center}
\caption{\em{The 24 SV40 capsid proteins located on the plane orthogonal to a 2-fold symmetry axis represented by the red dots, when considering the symmetry - corrected version pictured in Fig.~\ref{SVA1}(c).}}
\label{fig:icosaunmovedsv40}
\end{figure}

  {\bf {\em SV40:}} We argued this viral capsid has an approximate centre of inversion, and we thus perform our analysis with the full icosahedral group. Besides the identity, which leaves all 360 capsid  proteins unmoved, we learn from the symmetry-corrected tiling
  in Fig.~\ref{SVA1}(c) that there are 24 capsid proteins lying on the plane orthogonal to  a given 2-fold symmetry axis, as can be observed in Fig.~\ref{fig:icosaunmovedsv40}.

  Hence the coefficients $n_p$ are given by,
  \be
n_p^{SV40}=\frac{1}{120}\left\{1080 \,\chi^p(e)\,+\,360\,\chi^p(g_0g_2) \right\}, \quad p=1_{\pm}, 3_{\pm}, 3'_{\pm},  4_{\pm}, 5_{\pm}.
\ee
Once again, using the information in the character table~\ref{charTable}, we obtain the following decomposition of the displacement representation,
\be \label{decompsv40}
\Gamma^{displ\,'}_{1080,SV40}= 12\Gamma^1_+ + 24\Gamma^3_+ + 24\Gamma^{3'}_+ +36\Gamma^4_+ +
48\Gamma^5_+ 
+ 6\Gamma^1_- + 30\Gamma^3_- + 30\Gamma^{3'}_- + 36\Gamma^4_- +
42\Gamma^5_-.
\ee
The genuine modes of vibrations are encoded in the following,
 \be
\Gamma_{1080,SV40}^{vib}=  12\Gamma^1_+ + 23\Gamma^3_+ + 24\Gamma^{3'}_+ +36\Gamma^4_+ +
48\Gamma^5_+ 
+ 6\Gamma^1_- + 29\Gamma^3_- + 30\Gamma^{3'}_- + 36\Gamma^4_- +
42\Gamma^5_-,
\ee
while the Raman modes of vibration are
\be 
 \Gamma_{1080,SV40}^{Raman}= 12\Gamma^1_+ + 48\Gamma^5_+,
 \ee
 i.e. the Raman active fundamental levels are the twelve non-degenerate levels belonging
to $\Gamma^1_+$ and the forty-eight five-degenerate levels belonging to $\Gamma^5_+$.  These Raman modes are not infrared active as expected.

\vskip 1cm
 {\bf {\em HK97:}}  This viral capsid does not exhibit a centre of inversion, even approximate, and the relevant group for our analysis is  the subgroup ${\cal I}$ of the full icosahedral group. The only element which leaves proteins unmoved is the identity, under which all 420 proteins are unmoved. The coefficients of the decomposition of the displacement representation in this case are
 \be
n_p^{HK97}=\frac{1}{60}\left\{1260 \,\chi^p(e)) \right\}, \quad p=1_{+}, 3_{+}, 3'_{+},  4_{+}, 5_{+}.
\ee
 We therefore have the decomposition
 \be
 \Gamma^{displ\,'}_{1260,HK97}= 21\Gamma^1_+ + 63\Gamma^3_+ + 63\Gamma^{3'}_+ +84\Gamma^4_+ +
105\Gamma^5_+ ,
\ee
and once the rotations and translations are subtracted, we obtain
 \be
 \Gamma_{1260,HK97}^{vib}= 21\Gamma^1_+ + 61\Gamma^3_+ + 63\Gamma^{3'}_+ +84\Gamma^4_+ +
105\Gamma^5_+. 
\ee
The Raman modes of vibrations are
\be 
 \Gamma_{1260,HK97}^{Raman}= 21\Gamma^1_+ + 105\Gamma^5_+ ,
 \ee
 i.e. the Raman active fundamental levels are the twenty-one non-degenerate levels belonging
to $\Gamma^1_+$ and the one hundred and five five-degenerate levels belonging to $\Gamma^5_+$. These modes are not infrared active.
\section{Conclusion}

A full analysis of the vibrational modes of icosahedral viral capsids is a challenging task, as the number of degrees of freedom for a capsid consisting of $n$ atoms is $3n$. Even in the framework of classical mechanics, and within the approximation of a harmonic potential to describe interactions between atoms, a brute force calculation requires increasingly prohibitive CPU times as the capsid's size grows. There are a handful of impressive results in the literature where various levels of coarse-graining have been implemented, and where use of icosahedral symmetry has helped reducing the computer time needed to obtain the frequencies of the lowest  modes of vibration for several capsids up to triangulation number $T=7$ \cite{TamaBrooks, VanVlijmen}. Although these calculations are extremely valuable, they offer little insight in patterns of vibrations across the spectrum of capsids.

Our motivation has been to step back from these molecular dynamics calculations, and re-examine if the underlying group theory, combined with Viral Tiling Theory, could provide new insights in the vibration patterns of icosahedral capsids.  The first step in our approximation procedure has consisted in replacing each capsid protein by a point mass whose location coincides with that of the centre of mass of the protein in question, calculated from all crystallographically identified constituent atoms. This level of coarse-graining may appear very crude, but we have shown here that it highlights features of the effective distribution of proteins on the capsid, that are buried in the existing analyses. These features emerged when we  analyzed the deviation between the distribution of point masses described above and the distribution obtained by taking the (additive) inverse of all point masses. Although strictly speaking, none of the known icosahedral viral capsids  enjoys a centre of inversion in vivo, some are remarkably close to possessing one, as can be quantified from the measured deviations. We found that the largest deviation between the actual location of a centre of mass and its would-be location, were there an exact centre of inversion for Bacteriophage MS2 ($T=3$), occurs in chain A and is $5.77\AA$ for a capsid determined at a resolution of $2.8\AA$. Similarly, the largest deviation for Simian Virus 40 (pseudo $T=7d$) occurs for chain F at $7.64\AA$ for a capsid determined at a  resolution of $3.1\AA$. These deviations are substantially smaller than the corresponding ones in chain B of Tomato Bushy Stunt Virus ($T=3$) where the deviation is $10.03\AA$ at a resolution of $2.9\AA$, and in chain D of Bacteriophage Hong Kong 97, where it  is $17.84\AA$. At the light of these data, we have considered that MS2 and SV40 had effective centres of inversion, as can be anticipated from their ideal tiling representations in Fig.~\ref{fig:icosaidealrhombictiling} and Fig.~\ref{fig:sv40bis}. Indeed, for MS2, the ideal tiling has a centre of inversion, while the SV40, although  not possessing one, has decorations which are highly compatible with invariance under inversion, as the six chains in the kite-shaped fundamental domain (of the proper rotation subgroup ${\cal{I}}$) are nearly on the edges of the half-kite- shaped fundamental domain of the full icosahedral group $H_3$ represented in Fig.~\ref{SVA1}(c).  We therefore conclude that the underlying  tiling and its decorations can help identify those capsids with an effective centre of inversion, and hence provide a way to qualitatively differentiate between vibrational patterns, as the underlying group theory yields subtle variations, particularly in the number of normal modes which are Raman active. 

Our analysis clarifies that the property for a viral capsid of  having a centre of inversion is neither correlated with its size, nor with a type of tiling. 
For instance, MS2 and HK97 are both modelled by a rhomb tiling, but the former has a centre of inversion while the latter does not. The important factor is how the capsid proteins' centres of mass are distributed on the tiles within the fundamental domain of the icosahedral group.  As for the level of coarse-graining adopted here, we believe it provides a reasonable estimate of the group theoretical properties of the lowest modes of vibrations, and will yield useful information on the actual frequencies of these modes \cite{Peeters Taormina}.

The methods developed here can easily be used for the analysis of any known capsid, irrespective of its size, and should provide valuable information on its  lowest frequency modes of vibration.

\vskip 2cm

{\bf Acknowledgments}

AT is indebted to Kasper Peeters for sharing his computer expertise and providing stimulating discussions. She thanks EPSRC for a Springboard Fellowship EP/D062799 and the Departments of Biology  and Mathematics at York University for their hospitality. RT gratefully acknowledges an EPSRC Advanced Research Fellowship, and financial support for KME via the  Research Leadership Award F/00224/AE from the Leverhulme Trust.
\newpage
\section*{Appendix }
\renewcommand{\theequation}{A.\arabic{equation}}
  \setcounter{equation}{0}  
Consider  Fig.~\ref{fig:unmoved}(a) and (b). 
The 12 vertices of the icosahedron can be thought of as the vertices of three golden rectangles which are  perpendicular to each other.  
If we choose a right handed Cartesian reference frame in Fig.~\ref{fig:unmoved}(b), and scale the icosahedron in such a way that its edges have length 2, the three golden rectangles have vertices $[ABCD]$, $[EFGH]$ and $[IJKL]$ with the coordinates being
\be
\begin{array}{lll}
A=(0,-1,\tau),& E=(1,\tau,0),& I=(\tau,0,1),\\
B=(0,1,\tau),& F=(-1,\tau,0),& J=(\tau,0,-1),\\
C=(0,-1,-\tau),& G=(-1,-\tau,0),&K=(-\tau,0,1),\\
D=(0,1,-\tau),& H=(1,-\tau,0),& L=(-\tau,0,-1),
\end{array}\nonumber\ee
where $\tau=\hf(1+\sqrt{5})$.  

We choose the elements $g_2$ and $ g_3$ as follows (see also, for instance, \cite{Hoyle}):
\begin{enumerate}
\item $g_2$ is the 2-fold rotation taken clockwise about the axis through the origin and the midpoint of
the segment [CH], given by $\frac{1}{2}(1,-(\tau+1),-\tau)$. The unit vector along this axis is $\frac{1}{2}(-\tau',-\tau,-1)$.
\item $g_3$ is the 3-fold rotation taken clockwise about the axis through the origin and the centroid of
the triangle [CHJ], given by $\frac{1}{3}(\tau+1,-(\tau+1),-(\tau+1))$. The unit vector along this axis is $\frac{1}{3}(1,-1,-1)$.
\end{enumerate}
 The full icosahedral group $H_3$ has 120 elements and is generated by the elements $g_2, g_3$ and $g_{0}$. The first two are well-chosen 2-fold and 3-fold rotations satisfying $g_2^2=g_3^3=1$ and
$(g_2g_3)^5=1$. They generate the group of proper rotations $I$ of order 60.

To obtain $H_3$, one adds the reflection through the origin (or inversion) $g_{0}$ such that $g_{0}^2=1$.

The elements of $H_3$ are organized in ten conjugacy classes, and hence there exist ten irreducible representations whose characters are given in Table~\ref{charTable}. 
\begin{landscape}
\begin{table}[h!b!p!]
\begin{tabular}{c|ccccc|ccccc||c|c}
{\rm Conj. class}&${\cal C}(e)$&${\cal C}(g_5)$&${\cal C}(g_5^2)$&${\cal C}(g_3)$&${\cal C}(g_2)$&
${\cal C}(g_{0})$&${\cal C}(g_{0}g_5)$&${\cal C}(g_{0}g_5^2)$&${\cal C}(g_{0}g_3)$&${\cal C}(g_{0}g_2)$&{\rm vectors}&{\rm 2-rank tensors}\\
{\rm size}&1&12&12&20&15&1&12&12&20&15&&\\
\hline
{\em irrep}&&&&&&&&&&&&\\
$\Gamma^1_+$&1&1&1&1&1&1&1&1&1&1&&$x^2+y^2+z^2$\\
$\Gamma^3_+$&3&$\tau$&$\tau'$&0&-1&3&$\tau$&$\tau'$&0&-1&$(R_x,R_y,R_z)$&\\
$\Gamma^{3'}_+$&3&$\tau'$&$\tau$&0&-1&3&$\tau'$&$\tau$&0&-1&&\\
$\Gamma^4_+$&4&-1&-1&1&0&4&-1&-1&1&0&&\\
$\Gamma^5_+$&5&0&0&-1&1&5&0&0&-1&1&&$(2z^2-x^2-y^2,x^2-y^2,$\\
&&&&&&&&&&&&$xy,yz,xz)$\\
\cline{1-6}&&\\
\end{tabular}
\begin{tabular}{c|cccccccccc||c|c}
\phantom{Conj. class}&\phantom{C(e)}&\phantom{C(g5)}&\phantom{C (g)}& \phantom{C(g)}&\phantom{C(g2).}&
\phantom{C(g)}&\phantom{ C(g0)}&\phantom{C(g0)}&\phantom{C(g0g0)}&\phantom{C(g0g))}&\phantom{(Rx,Ry,Rz) }&\phantom{ 2-rank tensors}\\
$\Gamma^1_-$&1&1&1&1&1&-1&-1&-1&-1&-1&&\\
$\Gamma^3_-$&3&$\tau$&$\tau'$&0&-1&-3&$-\tau$&$-\tau'$&0&1&$(x,y,z)$&\\
$\Gamma^{3'}_-$&3&$\tau'$&$\tau$&0&-1&-3&$-\tau'$&$-\tau$&0&1&&\\
$\Gamma^4_-$&4&-1&-1&1&0&-4&1&1&-1&0&&\\
$\Gamma^5_-$&5&0&0&-1&1&-5&0&0&1&-1&&\\
&&&&&&&&&&&&\\
\end{tabular}
\caption{Character table for the icosahedral group $H_3$ with $\tau=\frac{1}{2}(1+\sqrt{5})$ and $\tau'=\frac{1}{2}(1-\sqrt{5})$. The top left highlighted table corresponds to the group $I$ of proper rotations. When considering $I$ only, $(x,y,z)$ is assigned to the representation $\Gamma^3_+$ (adapted from \cite{Cotton, Hoyle}). \newline \newline
{\em Link with conventions in the Chemistry literature}: the suffix $+$ is often replaced by $g$ for gerade (even), and the suffix $-$ replaced by $u$ for ungerade (odd). Furthermore, $\Gamma^1 \equiv A$, $\Gamma^3 \equiv T_1$, $\Gamma^{3'} \equiv T_2$, $\Gamma^4 \equiv G$ and $\Gamma^5 \equiv H$.}
\label{charTable}
\end{table}
\end{landscape}


\begin{thebibliography}{99}
\bibitem{Sherman} M. B. Sherman et al, {\em Removal of divalent cations induces structural transitions in Red Clover Necrotic Mosaic Virus, revealing a potential mechanism for RNA release}, J. Virol.  {\bf 80(21)} (2006) 10395Ð10406.
\bibitem{Orlova} E. V. Orlova, private communication.
\bibitem {TamaBrooks} F. Tama and C. L. Brooks, {\em Diversity and identity of mechanical properties of icosahedral viral capsids studied with elastic network normal mode analysis}, J. Mol. Biol. {\bf 345} (2005)  299Ð314.
\bibitem{VanVlijmen} H. W. T. van Vlijmen and M. Karplus, {\em Normal mode calculations of icosahedral viruses with full dihedral flexibility by use of molecular symmetry}, J. Mol. Biol. {\bf 350} (2005)  528Ð542.
\bibitem{Tirion} M. M. Tirion, {\em Large amplitude elastic motions in proteins from a single-parameter, atomic analysis},  Phys. Rev. Lett. {\bf 77} (1996) 1905-1908.
\bibitem{Simonson} T. Simonson and D. Perahia, {\em Normal modes of symmetric protein assemblies. Application to the tobacco mosaic virus protein disk}, Biophysical Journal {\bf 61} (1992) 410-427.
\bibitem{Weeks} D. E. Weeks and W. G. Harter, {\em Rotation-vibration spectra of icosahedral molecules. II. Icosahedral symmetry, vibrational eigenfrequencies, and normal modes of buckminsterfullerene}, J. of Chem. Phys. {\bf 90} (1989) 4744-4771; 
Z. C. Wu, D. A. Jelski and T. F. George, {\em Vibrational modes of Buckminster fullerenes},  Chem. Phys. Lett. {\bf 137} (1987) 291-294. 
\bibitem{James} G. D. James, {\em The representation theory of Buckminster fullerenes}, 
J. of Algebra {\bf 167} (1994) 803-820.
\bibitem{Nakamoto} K. Nakamoto and A. McKinney, {\em Application of the Correlation Method to Vibrational Spectra of C60 and Other Fullerenes: Predicting the Number of IR- and Raman-Active Bands}, J. Chem. Educ. {\bf 77} (2000) 775; F. Rioux, {\em Vibrational Analysis for C60 and Other Fullerenes}, J. Chem. Educ. {\bf 80} (2003) 1380. 
\bibitem{Twarock} R. Twarock, {\em A tiling approach to virus capsid assembly explaining a structural puzzle in virology}, J. Theor. Biol. {\bf 226} (2004) 477; {\em The architecture of viral capsids based on tiling theory}, J. Theor. Medicine {\bf 6} (2005) 87-90.
\bibitem{Peeters Taormina} K. Peeters and A. Taormina, in preparation.
\bibitem{Raman} C. V. Raman and K. S. Krishnan, {\em A new type of Secondary Radiation}, Nature, {\bf 121} (1928) 501.
\bibitem{Humphreys} J. E. Humphreys, {\em Reflection groups and Coxeter groups}, Cambridge Studies in Advanced Mathematics, {\bf 29}, Cambridge University Press, Cambridge, 1990. 
\bibitem{Patera}  J. Patera and R. Twarock,  {\em Affine extensions of noncrystallographic Coxeter groups and quasicrystals}, J. Phys. A, Math. Gen. {\rm 35} (2002) 1551.
\bibitem{TwarockKeef} T. Keef and R. Twarock, {\em A new series of polyhedra as blueprints for viral capsids in the family of Papovaviridae}, Arxiv q-bio.BM/0512047.
\bibitem{Cornwell} J. F. Cornwell, {\em ` Group theory in physics'}, v.1, Academic Press, 1984, UK; S. Sternberg, {\em Group theory  and physics}, Cambridge University Press, 1994.
\bibitem{Herzberg} G. Herzberg, {\em ` Molecular spectra and molecular structure. II: Infrared and Raman spectra of polyatomic molecules'}, Lancaster Press (Lancaster, PA), 1945;
D. M. Bishop, {\em Group theory and chemistry}, Dover Publications, New York, 1973. 
\bibitem{Cotton} F. A. Cotton, {\em Chemical Applications of Group Theory}, 3rd edition,
Wiley, 1963.
\bibitem{Hoyle} R. B. Hoyle, {\em Shapes and cycles arising at the steady bifurcation with icosahedral symmetry}, Physica D, 191, 261-281.
\end{thebibliography}
\end{document}